\documentclass[review,3p]{elsarticle}

\usepackage{lineno,hyperref}
\usepackage{gensymb}   % Enable to use \degree
\modulolinenumbers[5]

\usepackage{amssymb,amsmath}
\usepackage{multirow}
\usepackage{gensymb}
\usepackage{bm}   % gerer le gras
\usepackage{epstopdf}
\usepackage{graphicx}
\usepackage{float}
\usepackage{subfig}
\usepackage{supertabular}
\usepackage{verbatim}
\usepackage{array}
\usepackage{color}
\usepackage{makecell}
\usepackage{framed}
\usepackage{longtable,tabularx}
\usepackage{nomencl}
\usepackage{booktabs}
\usepackage{natbib}
\usepackage[export]{adjustbox}
\makenomenclature

\newcommand{\reffig}[1]{figure \ref{#1}}
\newcommand{\reffigs}[1]{figures \ref{#1}}
\newcommand{\refFig}[1]{Figure \ref{#1}}
\newcommand{\refFigs}[1]{Figures \ref{#1}}
\newcommand{\reftab}[1]{table \ref{#1}}

\newcommand{\refeq}[1]{Eq. \eqref{#1}}
\newcommand{\refse}[1]{section \ref{#1}}

\journal{Mechanical Systems and Signal Processing}

\begin{document}

\begin{frontmatter}

\title{Coupled dynamics of steady jet flow control for flexible membrane wings}%\tnoteref{mytitlenote}}

%\author{Gael Kemp, Guojun Li and Rajeev K. Jaiman \corref{mycorrespondingauthor}}
%\address{Mechanical Engineering Department, National University of Singapore, 119077, Singapore}
%\fntext[footnoteRajeev]{Since 1880.}

\author[mymainaddress,mysecondaryaddress]{Guojun Li}

\author[mythirdaddress]{Rajeev Kumar Jaiman}

\author[mymainaddress,mysecondaryaddress]{Hongzhong Liu\corref{mycorrespondingauthor1}}
\cortext[mycorrespondingauthor1]{Corresponding author}
\ead{hzliu@mail.xjtu.edu.cn}

\address[mymainaddress]{School of Mechanical Engineering, Xi'an Jiaotong University, Xi'an, Shaanxi, China 710048}

\address[mysecondaryaddress]{State Key Laboratory for Manufacturing Systems Engineering, Xi'an Jiaotong University, Xi'an, Shaanxi, China 710054}

\address[mythirdaddress]{Mechanical Engineering, University of British Columbia, Vancouver, BC Canada V6T 1Z4}

\begin{abstract}
We present a steady jet flow-based flow control of flexible membrane wings for an adaptive and efficient motion of bat-inspired drones in complex flight environments. A body-fitted variational computational aeroelastic framework is adopted for the modeling of fluid-structure interactions. High-momentum jet flows are injected from the leading edge and transported to the wake flows to alter the aerodynamic performance and the membrane vibration. The phase diagrams of the coupled fluid-membrane dynamics are constructed in the parameter space of the angle of attack and the jet momentum coefficient. The coupled dynamical effect of active jet flow control on the membrane performance is systematically explored.  While the results indicate that the current active flow control strategy performs well at low angles of attack, the effectiveness degrades at high angles of attack with large flow separation. To understand the coupling mechanism, the variations of the vortex patterns at different jet momentum coefficients are examined by the proper orthogonal decomposition modes in the Eulerian view and the fluid transport process is studied by the coherent flow structures in the Lagrange description. Two scaling relations that quantitatively connect the membrane deformation with the aerodynamic loads presented in our previous work are verified even when active jet flow control is applied. A unifying feedback loop that reveals the fluid-membrane coupling mechanism is proposed. This feedback loop provides useful guidance for designing optimal active flow control strategies and enhancing flight capabilities. These findings can facilitate the development of next-generation bio-inspired drones that incorporate smart sensing and intelligent control.
\end{abstract}


\end{frontmatter}

\section{Introduction}
% Background and motivation
Bats possess extraordinary flight skills in a wide range of complex flight environments. Active flow control applied to flexible membrane wings is the core mechanism for bats to enhance aerodynamic performance and achieve high maneuverability \cite{cheney2022bats}. The flexible membrane itself can adapt to different environments with good performance through structural deformation using the fluid-membrane coupling effect \cite{cheney2014membrane}. However, these passively deformed membrane wings have limited ability to respond to time-varying and complex flight environments \cite{de2022fsi,tiomkin2022unsteady}. Due to the limitation of the morphing capability, high-momentum jet flow transported into wake flows provides an effective and additional way to further improve flight performance. Morphing membrane structures incorporated with jet flow control inspired by the bat flight exhibit tantalizing potential in the design of the next-generation intelligent bio-inspired drones. The study of jet flow control strategies for the fluid-membrane coupling system remains relatively unexplored in the development of morphing wings with smart sensing and intelligent control.

% fluid-membrane interaction
Flexible membrane wings have been widely demonstrated for their successful application in morphing aircraft. The flexibility of the membrane wing plays the role of coordinator between aerodynamic loads and structural deformation, which is beneficial to increasing lift, reducing drag and alleviating gusty disturbance \cite{rojratsirikul2010unsteady,serrano2018fluid}. As revealed in Li et al.\cite{li2022aeroelastic,li2020flow_accept}, the aerodynamic performance of the membrane wing is mainly governed by two factors, namely the camber effect and the flow-induced vibration, respectively. The flexible membrane is driven by the aerodynamic loads to deform. In turn, the changed membrane profile can adjust the pressure distributions on its surface. The structural vibration is synchronized with the shedding vortex to excite particular aeroelastic modes by manipulating the natural frequencies. The feedback loop summarized in \cite{li2022aeroelastic} suggests that the selected fluid modes with particular vortex patterns can alter the fluid stress distributions and further govern the structural vibration through the flexibility effect. These passive flow and structure control manners exhibit obvious limitations when encountering extremely complex flight environments in large-flight spaces and wide-speed spaces. The active adaption to time-varying flight environments is lacking. Additional energy is necessary to inject into the coupled fluid-membrane system to adjust the flow features and the structural deformation characteristics.

% Active control for flexible membrane
Bats show smart and efficient flight postures during taking-off, landing, preying and avoiding predators. The secret of optimal bat flight is to actively change its wing profiles by skeletons and muscles after sensing the surrounding flow information via their tiny hairs \cite{sterbing2011bat}. Besides, the bat wings can introduce high-momentum fluid flows to the leeward surface through deflections and adjust the vortex patterns as needed. Inspired by the bat flight, some active control strategies are incorporated into the coupled fluid-membrane system to achieve optimal performance in multiple flight conditions and flight missions. For example, Breuer and his collaborators \cite{curet2014aerodynamic,bohnker2019control,bohnker2023integrated} demonstrated the integration of dielectric elastomer actuators into the flexible membrane to adjust the aerodynamic performance.  By applying varying voltages across the membrane, the membrane profile could be actively modified and the stall was delayed with  $20\%$ lift enhancement.  Buoso and Palacios \cite{buoso2015electro,buoso2017demand,buoso2017bat} numerically explored the closed-loop feedback control of flexible membrane wings by modeling the bat wing muscles via dielectric elastomers. Reduced-order models of the coupled fluid-membrane system were constructed based on the proper orthogonal decomposition method and coupled with the proportional–integral–derivative scheme to form the actively controlled system. The investigations indicated that the aerodynamic performance can be tuned by adjusting the actuating frequency. He et al. \cite{he2023aerodynamics} employed piezoelectric macro-fiber composite actuators onto the membrane wing to achieve open-loop active control. The lift enhancement and drag reduction were achieved simultaneously when the surrounding flows were stimulated by the actively vibrating membrane. An active flow control method for flexible membrane wings based on piezoelectric materials was numerically explored by Huang and his colleagues \cite{huang2021fluid,huang2022energy,xia2023mode} to achieve energy harvesting and vibration mode selection. The common technical points in the studies mentioned above are that the active control applied on the membrane wing is achieved by changing Young's modulus and the tension of the structure itself. However, the quantitative relationship between vortex patterns, structural deformation and aerodynamic loads remains unclear. Specifically, the actuation effect of structural deflection and vibration frequency on aerodynamic loads is still unknown.

% Steady jet flow control technology
In addition to the active control of the structural deformation, the active control of the fluid flows is another possible way to adjust the aerodynamic performance of the membrane wing. Recently, the integration of active jet flow control technology into morphing aircraft has become one of the important development trends in future advanced aircraft design, like the Control of Revolutionary Aircraft with Novel Effectors (CRANE) project launched by DARPA in 2020. The active jet flow control governs the coupled fluid-membrane system by directionally injecting additional momentum into the local flows near the wall. Therefore, the aerodynamic load distributions are affected due to the change of the vortex patterns based on the momentum equation \cite{li2021high}. As a result, the structural vibration can be suppressed through the fluid-structure coupling effect. The feasibility of active jet flow control technology has been widely verified in the wake control of rigid fixed-wing aircraft \cite{you2008active,itsariyapinyo2022experimental} and the surrounding flows around bluff bodies \cite{wang2016control,greco2020karman}. Jaiman and his collaborators demonstrated active jet control for the vortex-induced vibration of circular cylinders \cite{yao2017feedback} and spheres \cite{chizfahm2021data}. High-momentum jet flows are injected into the wake, shifting the unstable wake modes to the stable part. As a result, the synchronization between the vortex-shedding process and the structural vibration is altered, resulting in the suppression of vortex-induced vibration.

% Steady jet flow control for membrane     challenges
The integration of the active jet flow control into the coupled fluid-membrane system is still in the exploratory stage. Several challenges prevent the development of the active jet flow control in flexible membrane wings. Firstly, the flexible membrane inherently contains a wide spectrum of structural modes. The flow-induced vibration responses of the flexible membrane usually consist of multiple structural modes excited by the multi-scale and multi-modal fluid modes \cite{song2008aeromechanics,li2022aeroelastic}. As mentioned above, the vibration suppression for rigid bodies with low ($<$3) degrees of freedom can be achieved by shifting the vortex shedding frequency away from the natural frequency of the structure via active jet flow control. However, the flow-excited membrane instability is driven by several flow perturbation factors, like pressure fluctuations at low angles of attack \cite{serrano2018fluid}, shear layer instability at moderate angles of attack \cite{rojratsirikul2010unsteady} and vortex shedding at high angles of attack \cite{li2023unsteady}, rather than only vortex shedding for bluff bodies. Besides, the membrane instability can also be excited by the variation of the natural frequencies or the structural mode transition when the membrane is stretched under aerodynamic loads \cite{li2020flow_accept,xia2023mode}, which is different from the fixed natural frequency of rigid bodies. These complex mechanisms complicate the application of the active jet flow control to achieve vibration suppression and aerodynamic performance adjustment on flexible membrane wings. Then, different from the rigid wing, the wave-like moving boundary problem of the flexible membrane places an obstacle to developing the jet flow control. In particular, jet flow actuator placement strategies in the coupled fluid-membrane system should be explored. From a fundamental standpoint, it is important to study the parameter range when the jet-based flow control becomes effective and what is the underlying physical mechanism of jet-based flow control.

% FTLE   
Active jet flow control can directionally manipulate the fluid transport process and achieve aerodynamic performance adjustment. Understanding the relationship between fluid transport and vortex pattern evolution in the tight coupling system is the key to revealing the active flow control mechanism. The complex transport process of the near-wall fluid mixing with the injected high-momentum jet flows raises high demands on the transport analysis method for the unsteady flow scenario. The streamlines, the vorticity and pressure contours based on the Eulerian description just provide limited information at the examined time instant. This is mainly due to the time independence of the Eulerian description which excludes the dynamic properties. The well-known Lagrangian Coherent Structure (LCS) method based on the Lagrangian view proposed by Haller \cite{haller2001distinguished,haller2015lagrangian} offers an effective way to detect the hiding fluid skeletons and indicate the fluid transport boundary. The finite-time Lyapunov exponent (FTLE) method \cite{shadden2005definition} can extract the most attracting and repelling structures from the unsteady flow fields numerically and detect the LCSs by the ridges in the FTLE field. The FTLE method was widely used to visualize LCSs and analyze the fluid transport for flows past bluff bodies \cite{cao2021forced}, flapping airfoils \cite{chen2016using,wang2021study}, airfoils with active flow control \cite{cao2019lagrangian,cao2020lagrangian} etc. He and Wang \cite{he2020fluid} analyzed the effect of Reynolds numbers on the frequency characteristics of flexible membrane wings by visualizing the transient fluid fields by the FTLE ridges. The FTLE method was demonstrated as an effective way to bridge the gap between the fluid transport and the vortex pattern evolution.

% Feedback loop mechanism  remaining problems
The prerequisite for achieving optimal flight of the morphing aircraft like a bat is to fully explore the physical mechanism of fluid-membrane interaction. Plenty of efforts have been put in during the past decades to explore how the flexible membrane responds to the unsteady flow and adapts to complex flight environments with the aid of the active control techniques. Many previous studies remained somewhat qualitative descriptions of flow field evolution and structural deformation at different parameters \cite{song2008aeromechanics,rojratsirikul2011flow,sun2017nonlinear}. Some investigations have been performed to examine the instability of the coupled fluid-membrane system by linear stability analysis \cite{tiomkin2019membrane,mavroyiakoumou2021eigenmode}. 
In Li et al. \cite{li2022aeroelastic}, a feedback cycle between the vortex pattern, the pressure distribution and the membrane vibration was initially proposed. The excitation of particular vortex patterns and membrane vibration modes is attributed to the synchronization between the vortex shedding frequency and the membrane's natural frequency. Although some progress has been achieved in the study of the fluid-membrane coupling mechanism, a complete feedback loop that can link the key variables by the quantitative description based on simple physical relations to reveal the underlying mechanism is lacking. Moreover, the role of the steady jet flows injected into the feedback loop of fluid-membrane interaction could be examined for improved aerodynamic performance. This can open a new avenue and guideline for the design of next-generation morphing wings incorporating active flow control and achieving optimal flight performance.

% Our goal
In this manuscript, for the first time, we apply the active jet flow control to the coupled fluid-membrane system. The membrane aeroelasticity is numerically investigated by a recently developed three-dimensional partitioned aeroelastic framework. Of particular interest is to examine the effect of active jet flow control on the membrane dynamics and explore optimal active flow control strategies. With the aid of the FTLE method, the mode decomposition technique and the scaling relations, the following key questions concerning the fluid transport process and the active flow control mechanism are addressed: (i) How do active jet flows with high momentum affect the coupled fluid-membrane dynamics? (ii) Within what parameter range does active jet flow control work? (iii) How to quantitatively describe the relation between the vortex pattern, the aerodynamic loads and the membrane deformation? (iv) Are there any unifying feedback loops that reveal the underlying mechanisms of the coupled systems and general guidelines for the design of flexible membrane wings incorporating active flow control? To address (i), we apply steady jet flows on the membrane surface and perform a series of numerical simulations in the parameter space of the angle of attack and the momentum coefficient. The instantaneous flow features and the membrane responses are analyzed to examine the effect of the jet flow control. The variation of the performance for flexible membrane wings is compared at different parameters to answer (ii). We suggest two scaling relations to respond to question (iii). We further examine how the high-momentum jet flows are transported into the wake flows and change the vortex patterns with the aid of visualizing LCSs and POD modes. The vortex patterns are quantitatively connected with the aerodynamic loads through the momentum equations, which have been demonstrated in Li et al. \cite{li2021high}. The aeroelastic modes that link the structural deformation and the vortex pattern are excited via the frequency synchronization and the mode selection mechanism presented in \cite{li2022aeroelastic}. Based on these studies, a unifying feedback loop is proposed and the general guidelines are summarized.

% Organization
This paper is organized as follows: the computational membrane aeroelastic framework, the FTLE method and the POD technique are described in \refse{sec:section2}. We present the problem setup for the jet flow control applied to the flexible membrane in \refse{sec:section3}. In \refse{sec:section4}, the effect of the jet flow control, the coupled fluid-membrane dynamics and the underlying mechanism are examined in detail. Concluding remarks are provided in \refse{sec:section5}.

\section{Methodology} \label{sec:section2}
To simulate the coupled fluid–flexible structure system, the incompressible Navier–Stokes equations are solved together with the nonlinear structure equations via a partitioned iterative scheme. The Navier-Stokes equations are discretized via a stabilized Petrov–Galerkin finite element method in an arbitrary Lagrangian-Eulerian (ALE) reference frame. For the sake of completeness, we provide a brief review of the fluid-structure system and adopted data analysis tools based on the finite-time Lyapunov exponent and the proper orthogonal decomposition techniques.
\subsection{Computational membrane aeroelasticity}
The incompressible Navier-Stokes equations with the dynamic subgrid Large Eddy Simulation model in an arbitrary Lagrangian-Eulerian form are coupled with the flexible multibody structural motion equations to describe the coupled fluid-membrane system. The fluid domain $\Omega^f$ is discreted into $n_{el}^f$ non-overlapping finite elements $\Omega^e$ ($e=1,2,\cdots,n_{el}^f$) in space by a stabilized Petrov-Galerkin finite element method. We employ a generalized-$\alpha$ method \cite{jansen2000generalized} to calculate the fluid variables in the next time step $t^{n+1}$. A user-controlled spectral radius $\rho_{\infty}$ is employed to damp high-frequency errors. The updated fluid variables are given as
\begin{eqnarray}
	\hat{\boldsymbol{u}}^{f,n+1} = \hat{\boldsymbol{u}}^{f,n} + \Delta t \partial_t \hat{\boldsymbol{u}}^{f,n} + \gamma^f \Delta t (\partial_t \hat{\boldsymbol{u}}^{f,n+1}- \partial_t {\boldsymbol{u}}^{f,n}),
	\\
	\partial_t \hat{\boldsymbol{u}}^{f,n+\alpha_m^f} = \partial_t \hat{\boldsymbol{u}}^{f,n} + \alpha_m^f (\partial_t \hat{\boldsymbol{u}}^{f,n+1}- \partial_t \hat{\boldsymbol{u}}^{f,n}),
	\\
	\hat{\boldsymbol{u}}^{f,n+\alpha^f} = \hat{\boldsymbol{u}}^{f,n} + \alpha^f (\hat{\boldsymbol{u}}^{f,n+1}- \hat{\boldsymbol{u}}^{f,n}),
	\\
	\hat{\boldsymbol{u}}^{m,n+\alpha^f} = \hat{\boldsymbol{u}}^{m,n} + \alpha^f (\hat{\boldsymbol{u}}^{m,n+1}- \hat{\boldsymbol{u}}^{m,n}),
\end{eqnarray}
where $\Delta t$ is the time step size and $\partial_t$ denotes the partial derivative in time. The filtered fluid and mesh velocities at the time step $n$ are defined as $\hat{\boldsymbol{u}}^{f,n}$ and $\hat{\boldsymbol{u}}^{m,n}$, respectively. The generalized-$\alpha$ parameters $\alpha^f$, $\alpha_m^f$ and $\gamma^f$ that governed by the spectral radius $\rho_{\infty}$ are written as
\begin{equation}
	\alpha^f = \frac{1}{1+\rho_{\infty}}, \quad  \alpha_m^f = \frac{1}{2} \left( \frac{3-\rho_{\infty}}{1+\rho_{\infty}} \right), \quad  \gamma^f = \frac{1}{2} + \alpha_m^f - \alpha^f.
	\label{fem2}
\end{equation}

Suppose $\mathcal{S}_{\hat{\boldsymbol{u}}^f}^h$ and $\mathcal{S}_{{p}}^h$ are the trial function spaces for fluid velocity and pressure, which are defined as
\begin{eqnarray}
	\mathcal{S}_{\hat{\boldsymbol{u}}^f}^h = \{\hat{\boldsymbol{u}}^f_h | \hat{\boldsymbol{u}}^f_h \in H^1(\Omega^f(t)), \hat{\boldsymbol{u}}^f_h = \hat{\boldsymbol{u}}^f_D \ \text{on} \ \Gamma^f_D(t)\},
	\\
	\mathcal{S}_{{p}}^h = \{ {p}_h | {p}_h \in L^2(\Omega^f(t)) \},
\end{eqnarray}
where $H^1(\Omega^f(t))$ and $L^2(\Omega^f(t))$ represent the square-integrable $\mathbb{R}^d$-valued function space and the scalar-valued function space in the fluid domain $\Omega^f(t)$, respectively. The test function spaces for fluid velocity $\mathcal{V}^{h}_{\boldsymbol{\psi}^f}$ and for pressure $\mathcal{V}^{h}_q$ are given as
\begin{eqnarray}
	\mathcal{V}^{h}_{\boldsymbol{\psi}^f} = \{ \boldsymbol{\psi}^f_h | \boldsymbol{\psi}^f_h \in  H^1(\Omega^f(t)), \boldsymbol{\psi}^f_h = \boldsymbol{0} \ \text{on} \ \Gamma^f_D(t) \},
	\\
	\mathcal{V}^{h}_q = \{ q_h | q_h \in L^2(\Omega^f(t)) \},
\end{eqnarray}
where $\boldsymbol{\psi}^f_h$ and $q_h$ denote the weighting functions of fluid velocity $\hat{\boldsymbol{u}}^f_h$ and pressure ${p}_h$. 

The filtered incompressible Navier-Stokes equations in the variational statement are given as: find the velocity and pressure fields $[\hat{\boldsymbol{u}}^{f,n+\alpha^f}_h,\hat{p}_h^{n+1}] \in \mathcal{S}_{\hat{\boldsymbol{u}}^f}^h \times \mathcal{S}_{\hat{p}}^h$ such that $\forall[\boldsymbol{\psi}^f_h,q_h] \in \mathcal{V}^{h}_{\boldsymbol{\psi}^f} \times \mathcal{V}^{h}_q  $
\begin{eqnarray}
	\int_{\Omega^f(t^{n+1})} \rho^f (\partial_t \hat{\boldsymbol{u}}^{f,n+\alpha^f_m}_h \bigg|_{\boldsymbol{\chi}} + (\hat{\boldsymbol{u}}^{f,n+\alpha^f}_h - \boldsymbol{u}^{m,n+\alpha^f}_h) \cdot \nabla \hat{\boldsymbol{u}}^{f,n+\alpha^f}_h) \cdot \boldsymbol{\psi}^f_h {\rm{d}\Omega} \nonumber\\
	+\int_{\Omega^f(t^{n+1})} \hat{\boldsymbol{\sigma}}^{f,n+\alpha^f}_h:\nabla \boldsymbol{\psi}^f_h {\rm{d}\Omega} + \int_{\Omega^f(t^{n+1})} {\boldsymbol{\sigma}^{\text{sgs},n+\alpha^f}_h}:\nabla \boldsymbol{\psi}^f_h {\rm{d}\Omega} \nonumber\\
	+ \sum_{e=1}^{n^f_{el}} \int_{\Omega^e} \tau_m (\rho^f (\hat{\boldsymbol{u}}^{f,n+\alpha^f}_h-\boldsymbol{u}^{m,n+\alpha^f}_h) \cdot \nabla \boldsymbol{\psi}^f_h+\nabla q_h) \cdot \boldsymbol{\mathcal{R}}_m {\rm{d}\Omega^e} \nonumber\\
	-\int_{\Omega^f(t^{n+1})} q_h (\nabla \cdot \hat{\boldsymbol{u}}^{f,n+\alpha^f}_h) {\rm{d}\Omega}  +\sum_{e=1}^{n^f_{el}} \int_{\Omega^e} \nabla \cdot \boldsymbol{\psi}^f_h \tau_c \boldsymbol{\mathcal{R}}_c {\rm{d}\Omega^e} \nonumber\\
	-\sum_{e=1}^{n^f_{el}} \int_{\Omega^e} \tau_m \boldsymbol{\psi}^f_h \cdot (\boldsymbol{\mathcal{R}}_m \cdot \nabla \hat{\boldsymbol{u}}^{f,n+\alpha^f}_h) {\rm{d}\Omega^e} - \sum_{e=1}^{n^f_{el}} \int_{\Omega^e} \nabla \boldsymbol{\psi}^f_h : (\tau_m \boldsymbol{\mathcal{R}}_m \otimes \tau_m \boldsymbol{\mathcal{R}}_m) {\rm{d}\Omega^e} \nonumber\\
	= \int_{\Omega^f(t^{n+1})} \boldsymbol{b}^f(t^{n+\alpha^f}) \cdot \boldsymbol{\psi}^f_h {\rm{d}\Omega} + \int_{\Gamma^f_N} \boldsymbol{h}^f \cdot \boldsymbol{\psi}^f_h {\rm{d} \Gamma} ,
	\label{eq:LES1} 
\end{eqnarray}
where the terms in the first and second lines represent the Galerkin terms for the momentum equation and the viscous stress terms. Here, $\hat{\boldsymbol{\sigma}}^{f}_h$ denotes the Cauchy stress tensor and $\boldsymbol{\sigma}^{\text{sgs}}_h$ is the extra subgrid-scale stress term related to the subgrid filtering procedure in large eddy simulation. The integral of the Petrov-Galerkin stabilization terms for the momentum equation on each element domain is presented in the third line. In the fourth line, the Galerkin and the Galerkin/least-squares stabilization terms for the continuity equation are shown. The residual terms related to the approximation of the fine-scale velocity on element interiors are given in the fifth line. The two terms on the right-hand side of \refeq{eq:LES1} are the body forces and the Neumann boundary conditions. The element-wide residuals of the continuity and momentum equations are defined as $\boldsymbol{\mathcal{R}}_c$ and $\boldsymbol{\mathcal{R}}_m$, respectively. The stabilization parameters for the continuity and momentum equations are represented by $\tau_c$ and $\tau_m$.

Suppose the trial solution and the test function spaces are defined as ${\mathcal{S}_{\bm{u}^s}^s}$ and $\mathcal{V}^{s}_{\boldsymbol{\phi}^s}$
\begin{eqnarray}
	{\mathcal{S}_{\bm{u}^s}^s} = \{\bm{u}^s | \bm{u}^s \in H^1(\Omega^s(t)), \bm{u}^s = \bm{u}^s_d \ \text{on} \ \Gamma^s_d(t)\}
	\\
	\mathcal{V}^{s}_{\boldsymbol{\phi}^s} = \{ \boldsymbol{\phi}^s | \boldsymbol{\phi}^s \in  H^1(\Omega^s(t)), \boldsymbol{\phi}^s = \bm{0} \ \text{on} \ \Gamma^s_d(t) \}
\end{eqnarray}
where $\bm{d}^s$ and $\bm{u}^s$ represents the structural displacement and velocity, respectively. The square-integrable $\mathbb{R}^d$-valued function space and the scalar-valued function space in the structural domain $\Omega^s(t)$ are defined as $H^1(\Omega^s(t))$ and $L^2(\Omega^s(t))$. $\bm{u}^s_d $ is the structural velocity on the Dirichlet boundary $\Gamma^s_d(t)$ and $\boldsymbol{\phi}^s$ denotes the corresponding weighting-function counterpart. The variational formulation of the structural motion equations is given as: find $\bm{u}^s \in  {\mathcal{S}_{\bm{u}^s}^s}$ such that $\forall \boldsymbol{\phi}^s \in \mathcal{V}^{s}_{\boldsymbol{\phi}^s} $
\begin{eqnarray}
	&&\int^{t^{n+1}}_{t^n} 
	\left( \int_{\Omega^s_i} \rho^s \frac{\partial \bm{u}^s}{\partial t} \cdot \boldsymbol{\phi}^s {\rm{d}\Omega} + \int_{\Omega^s_i} \boldsymbol{\sigma}^s : \nabla \boldsymbol{\phi}^s  {\rm{d}\Omega}
	\right) {\rm{d}}t \nonumber\\
	&&= \int^{t^{n+1}}_{t^n} 
	\left(
	\int_{\Omega^s_i} \bm{b}^s \cdot \boldsymbol{\phi}^s  {\rm{d}\Omega} + \int_{\Gamma^s_i} \bm{h}^s \cdot \boldsymbol{\phi}^s {\rm{d} \Gamma}
	\right) {\rm{d}}t   
	\label{eq:eqMB1} 
\end{eqnarray}
where $\rho^s$ is the structural density and $\boldsymbol{\sigma}^s$ denotes the stress tensor. Herein, $\bm{h}^s=\boldsymbol{\sigma}^s \cdot \bm{n}^s$ represents the Neumann condition at the boundary $\Gamma^s_{i}$. $\bm{b}^s$ is the body force employing on the structure on the right-hand side of Eq.~(\ref{eq:eqMB1}).

On the fluid-structure interface $\Gamma^{fs}_{i}$, the velocity and traction continuity conditions must be satisfied for the coupled equations, which are given as
\begin{eqnarray}
	\hat{\bm{u}}^f(\boldsymbol{\varphi}^s(\bm{x}^s,t),t)=\bm{u}^s(\bm{x}^s,t) \quad \forall \bm{x}^s \in \Gamma^{fs}_i  
	\label{eq:eqFSI1}
	\\
	\int_{\boldsymbol{\varphi}^s(\gamma^{fs},t)}\hat{\boldsymbol{\sigma}}^f(\bm{x}^f,t) \cdot \bm{n}^f {\rm{d}}\Gamma+\int_{\gamma^{fs}}\boldsymbol{\sigma}^s(\bm{x}^s,t) \cdot \bm{n}^s {\rm{d}}\Gamma=0  \quad \forall \gamma^{fs} \in \Gamma^{fs}_i
	\label{eq:eqFSI2}
\end{eqnarray}
where $\boldsymbol{\varphi}^s$ is a function that maps the information at the structural point $\bm{x}^s$ to the deformed point in the fluid domain. The outer normals to the fluid-structure interface in the fluid and structural domains are defined as $\bm{n}^f$ and $\bm{n}^s$, respectively. $\boldsymbol{\varphi}^s(\gamma^{fs},t)$ denotes the fluid domain of the part $\gamma^{fs}$ at time instant $t$.
Owing to the body-fitted Lagrangian-Eulerian coupling of the fluid-structure system, the Dirichlet boundary condition is employed on the jet slot patch. Uniform flows with a fixed velocity $\boldsymbol{u}^{j}$ are imposed on the boundary $\Gamma^j$ in the fluid domain
\begin{eqnarray}
	\hat{\boldsymbol{u}}^{f} = \boldsymbol{u}^j  \quad \forall \bm{x}^f \in \Gamma^{j}
	\label{eq:eqjet}
\end{eqnarray}

\begin{figure}[H]
	\centering
	\includegraphics[width=0.8 \textwidth]{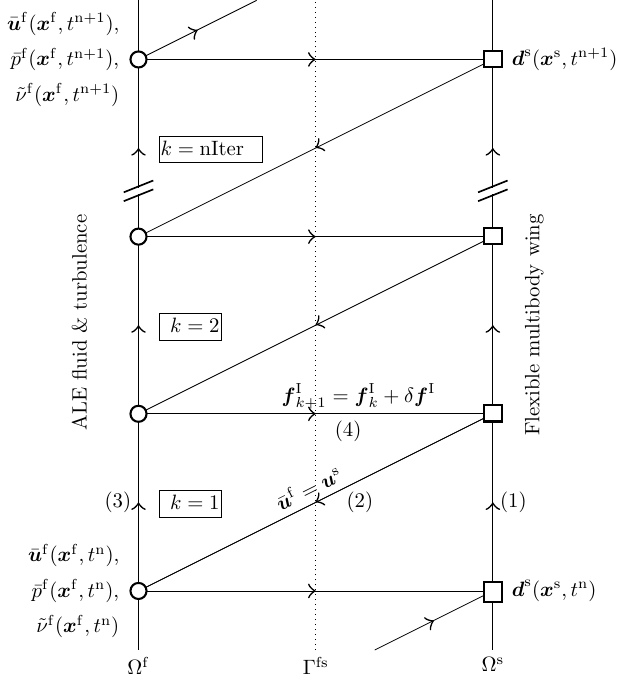}
	\caption{\label{fig:NIFC} A schematic of the predictor-corrector procedure for the coupled ALE fluid and flexible body through nonlinear iterative force correction scheme.}
\end{figure}

The incompressible Navier-Stokes equations and the flexible multibody structure equations are solved in a partitioned iterative manner \cite{li2018novel}. We employ a predictor-corrector approach to update the fluid and structure variables in time. The compactly-supported radius basis function (RBF) is utilized to exchange the aerodynamic forces and the structural displacements along the interface. The fluid mesh in a body-fitted manner is updated in the space based on the efficient RBF remeshing method. A high mesh quality can be preserved in a relatively large mesh motion condition. We implement the recently developed nonlinear interface force correction (NIFC) scheme \cite{jaiman2016stable} in the coupled aeroelastic framework. The purpose is to ensure numerical stability when the significant added mass effect occurs. The coupling algorithm based on the NIFC scheme is presented in \reffig{fig:NIFC}. This computational aeroelastic framework has been widely validated for flexible flapping wings \cite{li2018novel,li2021high} and morphing membrane wings \cite{li2020flow_accept}.

\subsection{Finite-time Lyapunov exponent}
The Lagrangian coherence structures can be approximated by the finite-time Lyapunov exponent fields which reflect the local stretching rate of the fluid flow. The ridges of the FTLE field are used to represent the stable and unstable manifolds, which indicate the boundaries of different dynamic parts in the flow field. The FTLE offers an effective way to visualize the fluid transport process from the Lagrangian view. In this study, the high-momentum jet flows are injected into the wake and mixed with the low-momentum flows near the boundary layer. Understanding these complex fluid transport phenomena with the aid of the FTLE method can help in revealing the active flow control mechanism.

The FTLE fields can be calculated from the stretching between two neighboring particles \cite{shadden2006lagrangian}. Considering the velocity field $\boldsymbol{u}^f(\boldsymbol{x}^f,t)$ in an Eulerian coordinate as a dynamic system from the Lagrangian viewpoint
\begin{eqnarray}
	\boldsymbol{\dot{x}}^f(t;t_0,\boldsymbol{x}_0^f) = \boldsymbol{u}^f(\boldsymbol{x}^f,t),  \quad \quad  \forall \boldsymbol{x}^f \in \Omega^f
\end{eqnarray}
where $\boldsymbol{{x}}^f(t;t_0,\boldsymbol{x}_0^f)$ denotes an arbitrary fluid particle at its current time $t$ released from the initial point $\boldsymbol{x}_0^f$ at time $t_0$ in the Lagrangian description. A flow map that maps fluid particles from their initial positions at time $t_0$ to the current positions at $t$ can be defined as
\begin{eqnarray}
	\boldsymbol{\varphi}_{t_0}^t = \boldsymbol{{x}}^f(t;t_0,\boldsymbol{x}_0^f) ,  \quad \quad  \forall \boldsymbol{x}^f \in \Omega^f
\end{eqnarray}

Therefore, the deformation of the flow near $\boldsymbol{x}_0^f$ within time interval $\left[ t_0,t \right]$ can be described by the flow gradient $\nabla \boldsymbol{\varphi}_{t_0}^t (\boldsymbol{x}_0^f)$. The right Cauchy-Green strain that represents the Lagrangian strain in the velocity field can be given as
\begin{eqnarray}
    \boldsymbol{C}_{t_0}^t (\boldsymbol{x}_0^f) = \left[ \nabla \boldsymbol{\varphi}_{t_0}^t (\boldsymbol{x}_0^f)  \right]^T \nabla \boldsymbol{\varphi}_{t_0}^t (\boldsymbol{x}_0^f)
\end{eqnarray}
where $\boldsymbol{C}_{t_0}^t (\boldsymbol{x}_0^f)$ is a $2 \times 2$ symmetric positive definite matrix for two-dimensional flows. The real positive eigenvalues and orthogonal real eigenvalues can be calculated from 
\begin{eqnarray}
	\boldsymbol{C}_{t_0}^t (\boldsymbol{x}_0^f) \boldsymbol{\xi}_i = \lambda_i \boldsymbol{\xi}_i, \quad \quad i=1,2
\end{eqnarray}
The FTLE field within the time interval $\left[ t_0,t \right]$ can be expressed by the maximum eigenvalue $\lambda_{max}$ of $\boldsymbol{C}_{t_0}^t (\boldsymbol{x}_0^f)$ as
\begin{eqnarray}
	\boldsymbol{\sigma}_{t_0}^t (\boldsymbol{x}_0^f) = \frac{1}{|t-t_0|} \ln \sqrt{\lambda_{\text{max}}(\boldsymbol{C}_{t_0}^t (\boldsymbol{x}_0^f))}
\end{eqnarray}

\begin{figure}[H]
	\centering
	\includegraphics[width=1.0 \textwidth]{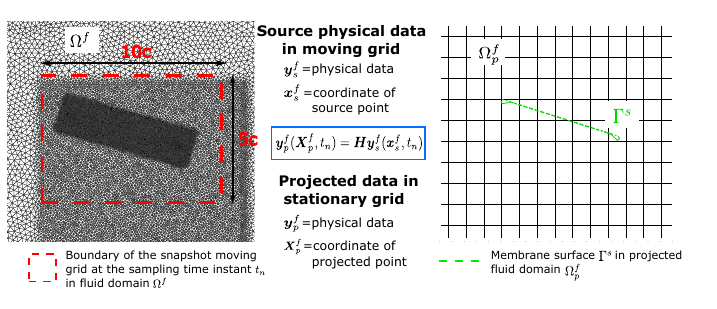}
	\caption{\label{fig:rbf} Illustration of flow variable projection from moving grid to stationary grid via the radial basis function method.}
\end{figure}

The LCSs in the flow field can be visualized as the ridges of the FTLE field $\boldsymbol{\sigma}_{t_0}^t (\boldsymbol{x}_0^f)$. The calculation of the FTLE field is to integrate the velocity field at fixed mesh points. Because of the mesh motion in the body-fitted mesh scheme, the velocity information at the moving mesh points is projected to a uniform Cartesian mesh with fixed positions by the mapping matrix $\boldsymbol{H}$ via the radial basis function method \cite{powell1985radial,li2022aeroelastic} as illustrated in \reffig{fig:rbf}. The projected velocity field obtained from numerical simulations is used to calculate the forward-time integration ($t>t_0$) and obtain the repelling LCSs at time $t_0$. Similarly, the attracting LCSs at time $t_0$ are obtained from the backward-time integration ($t<t_0$). The repelling LCSs contain the flow future information corresponding to the finite time stable manifolds, while the attracting LCSs include the flow historical information corresponding to the finite time unstable manifolds. These finite-time stable and unstable manifolds represent the boundaries between trajectories of fluid particles with different characteristics. Thus, the LCSs detected by the FTLE method are important to analyze the transport and mix process of the high-momentum jet flows in the wake of the flexible membrane wings.

\subsection{Proper Orthogonal Decomposition}  \label{sec:section2.3}
\begin{figure}[H]
	\centering
	\includegraphics[width=1.0 \textwidth]{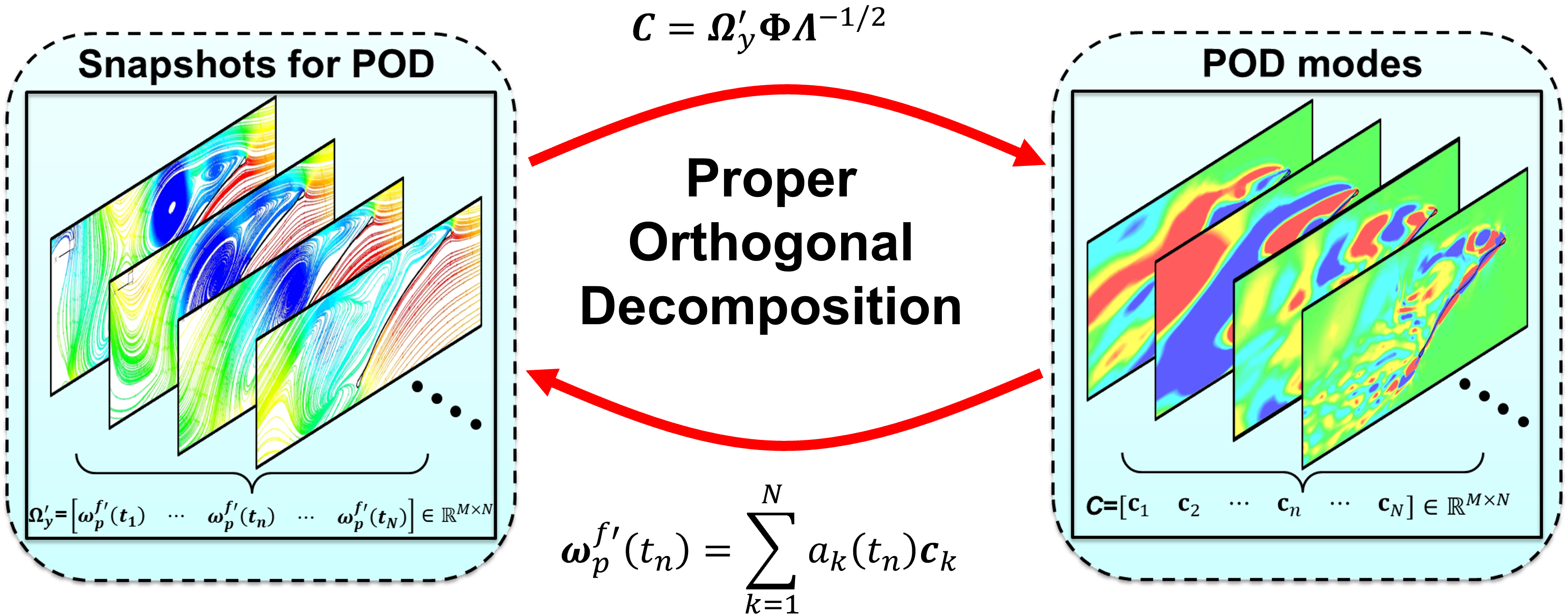}
	\caption{\label{fig:PODmode} A schematic of POD decomposition procedure for separated flow field with aeroelastic effects.}
\end{figure}

The vortex pattern in the coupled fluid-membrane system exhibits complex behaviors. This phenomenon is exacerbated by the transport and mixing of high-momentum jet flows in wakes. It is important to extract the coherent structures from the time-varying flow field and analyze the evolution of the vortex patterns. The proper orthogonal decomposition (POD) method \cite{lumley1970stochastic,sirovich1987turbulence} extracts the most energetic modes optimally from the flow field by diagonalizing the spatial correlation matrix. Different from the coherent structures detected by the FTLE ridges in the Lagrangian view, the POD modes show the coherent structures independent of time in the Eulerian coordinate. The POD modes provide a straight understanding of the most energetic coherent structures in the coupled system, which exclude the information of flow historical future information. \refFig{fig:PODmode} briefly presents the POD mode extraction procedure from the time-varying vorticity fields. The combined analysis of the LCSs and the POD modes for the coupled system can help in gaining a comprehensive understanding of the evolution of the vortex pattern when active jet flow control is applied.

Similar to the calculation of LCSs, the physical variables of interest for POD mode extraction should be projected from the moving mesh to a stationary grid before proceeding to mode decomposition. In this study, we focus on analyzing the POD modes from the time-varying vorticity fields. As illustrated in \reffig{fig:rbf}, the vorticity data at each moving grid is mapped to the Cartesian grid as
\begin{eqnarray}
	\boldsymbol{\omega}^f_p (\boldsymbol{X}^f_p,t_n) = \boldsymbol{H} \boldsymbol{\omega}^f_s (\boldsymbol{x}^f_s,t_n) \in \mathbb{R}^{M \times 1}
	\label{pod1}
\end{eqnarray}
where ${\boldsymbol{\omega}^f}_p (\boldsymbol{X}^f_p,t_n)$ denotes the projected vorticity data at $M$ discrete points. The fluctuating components are obtained by subtracting the mean values from the time-varying data. The collected snapshot fluctuating vorticity components at $N$ time instants is stored in a matrix form
\begin{eqnarray}
	\boldsymbol{\Omega}_y^{\prime} =  [ {\boldsymbol{\omega}^f_p}^{\prime}(t_1) \quad  {\boldsymbol{\omega}^f_p}^{\prime}(t_2)  \ldots {\boldsymbol{\omega}^f_p}^{\prime}(t_n)  \ldots  {\boldsymbol{\omega}^f_p}^{\prime}(t_N) ] \in \mathbb{R}^{M \times N}
	\label{pod2}
\end{eqnarray}

The snapshot covariance matrix $\boldsymbol{R}$ can be defined as
\begin{eqnarray}
	\boldsymbol{R} =\frac{1}{N}(\boldsymbol{\Omega}_y^{\prime})^T \boldsymbol{\Omega}_y^{\prime} \in \mathbb{R}^{N \times N}
	\label{pod3}
\end{eqnarray}

The eigenvectors $\boldsymbol{\Phi}$ and eigenvalues of $\boldsymbol{\Lambda}$ are calculated from
\begin{eqnarray}
	\boldsymbol{R} \boldsymbol{\Phi} = \boldsymbol{\Lambda} \boldsymbol{\Phi}
	\label{pod4}
\end{eqnarray}

The $N$ energetic POD modes calculated based on the snapshot POD method can be given as
\begin{eqnarray}
	\boldsymbol{C} = [\boldsymbol{c_1} \quad \boldsymbol{c_2} \ldots \boldsymbol{c_n} \ldots \boldsymbol{c_N}] = \boldsymbol{\Omega}_y^{\prime} \boldsymbol{\Phi} \boldsymbol{\Lambda}^{-1/2} \in \mathbb{R}^{M \times N}
	\label{pod5}
\end{eqnarray}

The time coefficients are written as
 \begin{eqnarray}
 	\boldsymbol{A} =  \boldsymbol{\Phi} \boldsymbol{\Lambda}^{-1/2}   \in \mathbb{R}^{N \times N}
 	\label{pod6}
 \end{eqnarray}

The vorticity field can be reconstructed from the POD modes by multiplying with corresponding mode coefficients
\begin{eqnarray}
   \boldsymbol{\Omega}_y^{\prime} =	\boldsymbol{C}  \boldsymbol{A}^T  \in \mathbb{R}^{M \times N}
   \label{pod7}
\end{eqnarray}

In this study, 1024 vorticity snapshots are extracted from the simulation results with a sampling frequency of $f_s$=125 Hz for extracting the POD modes. Before proceeding to mode decomposition, the vorticity data at the moving grid is mapped to a $10c \times 5c$ stationary grid based on \refeq{pod1}. Then, the collected data is formed in a matrix form in \refeq{pod2}. The snapshot covariance matrix $\boldsymbol{R}$ is calculated and the POD modes can be obtained from \refeq{pod5}. These POD modes can be used to analyze changes in coherence structures when different active jet controls are applied.

\section{Problem description} \label{sec:section3}
\begin{figure}[H]
	\centering
	\includegraphics[width=1.0 \textwidth]{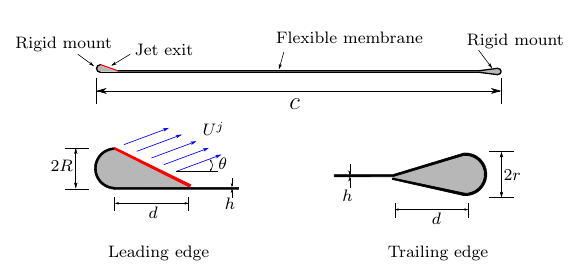}
	\caption{\label{fig:schematic} Schematic of flexible membrane wing geometry. Close-up view of geometry details of the rigid mounts at the leading and trailing edges.}
\end{figure}

A two-dimensional flexible membrane similar to the model established in the experimental study \cite{rojratsirikul2010unsteady} is considered in our numerical simulations. This membrane model is made of a black rubber sheet and it is attached to the rigid mounts at the leading and trailing edges. A schematic of the membrane wing geometry is shown in \reffig{fig:schematic}. The chord length of the membrane is $c$=150 mm and the thickness is $h$=0.2 mm. The rigid mount at the leading edge consists of a semicircle part with a diameter of $2R$=2.2 mm and a triangle part with a length of $d$=5.4 mm. The upper surface of the right-angled triangle part colored in red as shown in \reffig{fig:schematic} is the exit boundary of the jet flow. The rigid mount at the trailing edge is made of a semicircle part with a diameter of $2r$=1.5 mm and an isosceles triangle with a length of $d$=5.4mm. Uniform steady jet flows with a velocity of $U^j$ are imposed at the jet exit boundary to perform active flow control. The angle between the jet flows and the undeformed membrane is fixed at $\theta$=$20^\circ$. This angle ensures that jet flows with high momentum can be injected into the wake along the deformed membrane surface. The membrane is immersed in a uniform flow with a velocity of $U_{\infty}$=0.2886 m/s and the air density of $\rho^f$=1.1767 kg/m$^3$. The Young's modulus of the membrane is $E^s$=3346 Pa and the structural density is $\rho^s$=473 kg/m$^3$. Three non-dimensional parameters, namely Reynolds number, mass ratio and aeroelastic number, that govern the membrane dynamics are defined as
\begin{eqnarray}
	Re=\frac{\rho^f U_{\infty} c}{\mu}=2500, \quad  m^*=\frac{\rho^s h}{\rho^f c}=0.589,  \quad Ae=\frac{E^s h}{\frac{1}{2} \rho^f U^2_{\infty} c}=100.04
	\label{eq:parameter}
\end{eqnarray}

\begin{figure}[H]
	\centering
	\includegraphics[width=1.0 \textwidth]{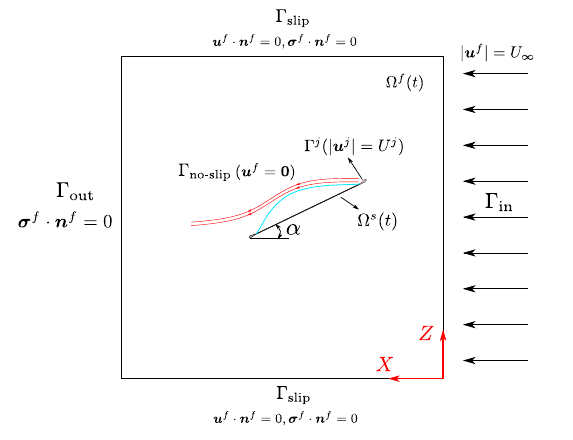}
	\caption{\label{fig:schematic2} Schematic of the computational domain for a two-dimensional membrane wing with active jet flow control immersed in a uniform fluid flow.}
\end{figure}

The flexible membrane is placed in the uniform freestream with an angle of attack of $\alpha$. Both leading and trailing edges are fixed and the flexible membrane is allowed to deform. A two-dimensional computational fluid domain shown in \reffig{fig:schematic2} is constructed for numerical simulations. The length of the computational domain is $40c\times40c$. We apply a uniform freestream velocity $U_{\infty}$ at the inlet boundary $\Gamma_{\text{in}}$. The traction-free boundary condition is employed along the outlet boundary $\Gamma_{\text{out}}$. The slip-wall boundary condition is applied at the up and bottom sides of the computational domain. The boundary condition on the membrane surface is set to the no-slip wall condition. A Dirichlet velocity condition is applied along the exit boundary to inject high-momentum jet flows into the wake.

The computational fluid domain is discretized by unstructured triangular elements. The boundary layer mesh is generated in a stretching ratio of 1.15 and to maintain $y^+<$1. The structured four-node rectangular finite element is utilized to discrete the structure model. The rigid mounts at the leading and trailing edges are modeled by the rigid body element. The geometrically exact co-rotational thin shell elements are employed to simulate the flexible membrane part. The flexible membrane is clamped at the leading and trailing edges with no pretension. The non-dimensional time step size is set to $\Delta t U_{\infty}/c$=0.00423 in the numerical simulations. 

We perform a mesh convergence study to select an appropriate mesh resolution. Three different mesh sets, namely M1, M2 and M3, are constructed for comparison purposes. A detailed comparison of the mesh convergence is presented in Appendix A. Before proceeding to investigate the active flow control effect, a detailed comparative study and validation are performed for the employed high-fidelity computational aeroelastic simulation, which is shown in Appendix B. Additional validations for jet flow control can be found in previous studies \cite{yao2017feedback,chizfahm2021data}.

The mean membrane deformation, the forces acting on the membrane wing and the pressure distribution on the wing surface are evaluated from the simulation data. The air load is calculated by integrating the surface traction taking into account the first layer of elements on the membrane surface. The instantaneous lift, the drag and the normal force coefficients are defined as
\begin{equation}
	C_L = \frac{1}{\frac{1}{2} \rho^f U_{\infty}^2 S} \int_{\Gamma} (\boldsymbol{\hat{\sigma}}^f \cdot \boldsymbol{e}) \cdot \boldsymbol{e}_z \rm{d \Gamma^s}
	\label{CL_2d} 
\end{equation}
\begin{equation}
	C_D = \frac{1}{\frac{1}{2} \rho^f U_{\infty}^2 S} \int_{\Gamma} (\boldsymbol{\hat{\sigma}}^f \cdot \boldsymbol{e}) \cdot \boldsymbol{e}_x \rm{d \Gamma^s}
	\label{CD_2d} 
\end{equation}
\begin{equation}
	C_n = \frac{1}{\frac{1}{2} \rho^f U_{\infty}^2 S} \int_{\Gamma} (\boldsymbol{\hat{\sigma}}^f \cdot \boldsymbol{e}) \cdot \boldsymbol{e}_c \rm{d \Gamma^s}
	\label{CN_2d} 
\end{equation}
where $\boldsymbol{e}_x$ and $\boldsymbol{e}_z$ are the Cartesian components of the unit normal $\boldsymbol{e}$ to the membrane surface and $\boldsymbol{e}_c$ is the unit normal to the chord line. $\boldsymbol{\hat{\sigma}}^f$ is the fluid stress tensor with $\Gamma^f$ being the surface boundary of the membrane. The pressure coefficient is defined as
\begin{equation}
	C_p = \frac{p-p_\infty}{\frac{1}{2} \rho^f U_{\infty}^2}
	\label{Cp_2d} 
\end{equation}
where $p$ and $p_\infty$ are the pressure at the concerned point and the pressure at the far-field, respectively. The momentum coefficient with a jet velocity of $U^j$ applying at the jet slot exit boundary $\Gamma^j$ is defined as
\begin{equation}
	C_{\mu} = \frac{\dot{m} U^{j}}{q_{\infty} S} = \frac{\rho^j L^j (U^j)^2}{\frac{1}{2}\rho^f c U_{\infty}^2}
	\label{Cu} 
\end{equation}
where $\dot{m}$ is the mass flow rate through the jet slot exit and $q_{\infty}$ represents the dynamic pressure of the freestream. The effective length of the jet exit boundary is the length normal to the jet flow direction, which is set to $L^j$=3.54 mm in the present study.

\section{Results and discussion} \label{sec:section4}
In the current study, we apply a steady jet flow control at the leading edge of the membrane to investigate the membrane aeroelastic dynamics. The effects of the momentum coefficient on the coupled fluid-membrane dynamics at different angles of attack are examined. We further investigate the mechanisms of lift improvement and structural vibration suppression. Finally, a unifying feedback loop and guidelines for active jet flow control are suggested.

\subsection{Coupled fluid-membrane dynamics} \label{sec:section4.1}
\begin{figure}[H]
	\centering
	\subfloat[]{\includegraphics[width=0.5 \textwidth]{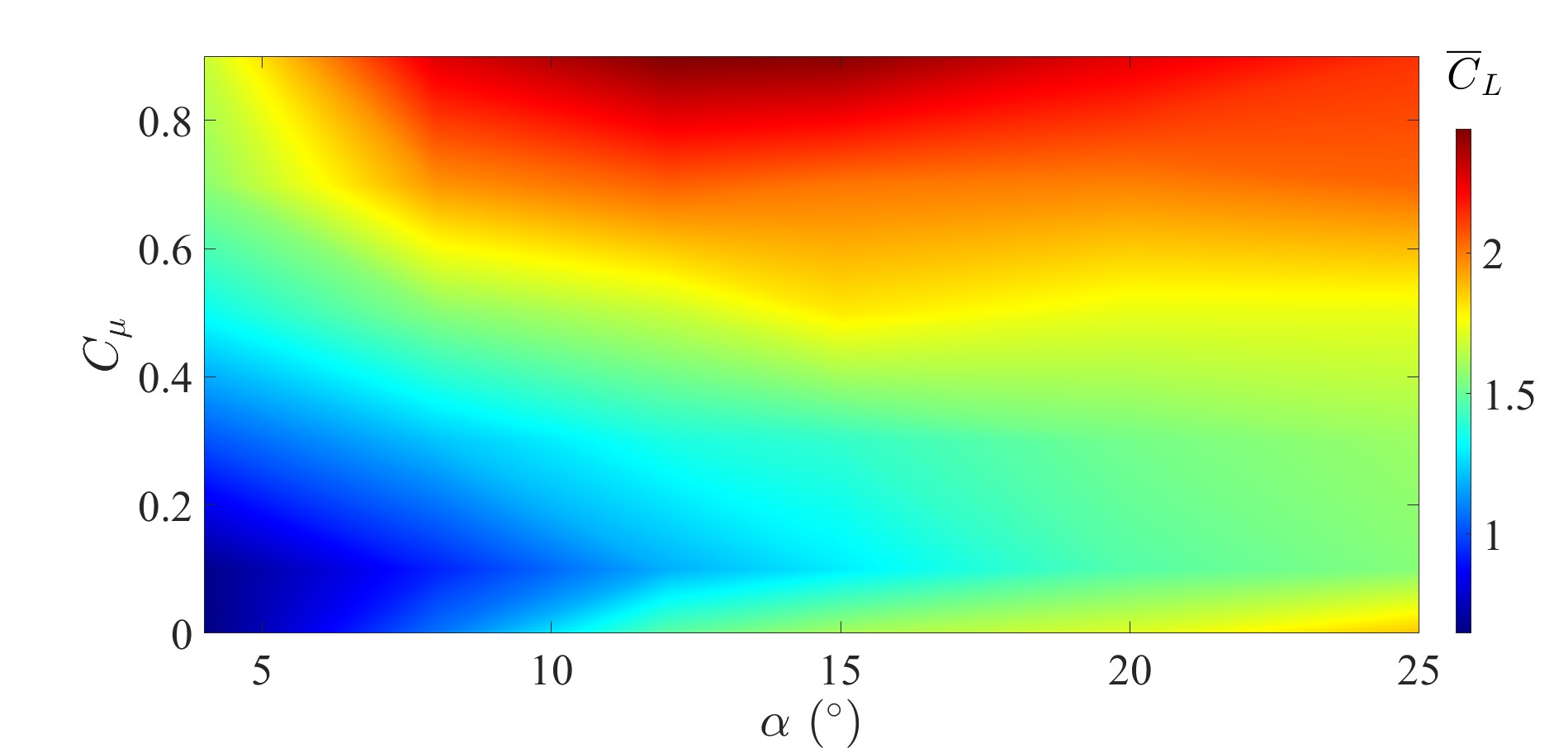}\label{fig:performancea}}
	\subfloat[]{\includegraphics[width=0.5 \textwidth]{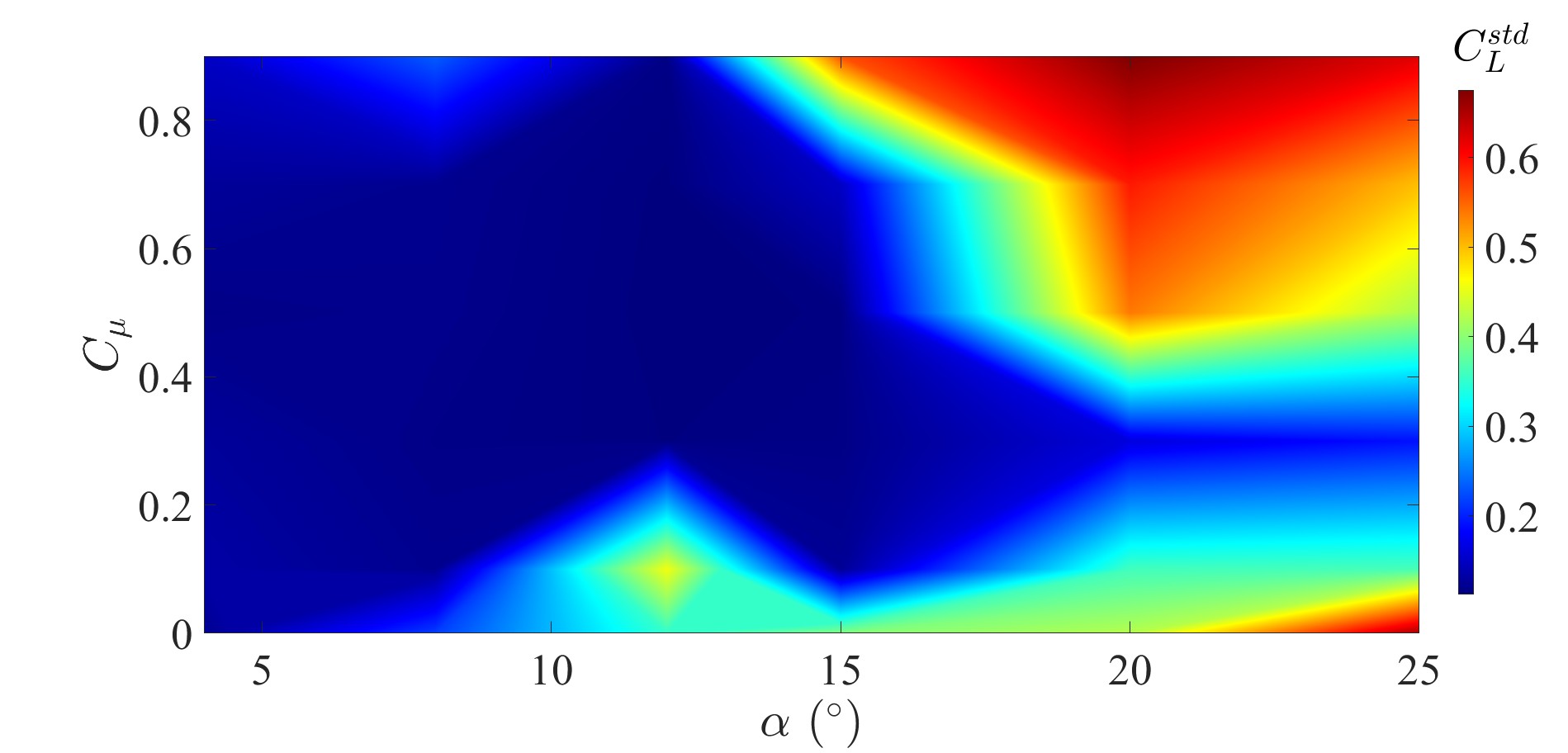}\label{fig:performanceb}}
	\\
	\subfloat[]{\includegraphics[width=0.5 \textwidth]{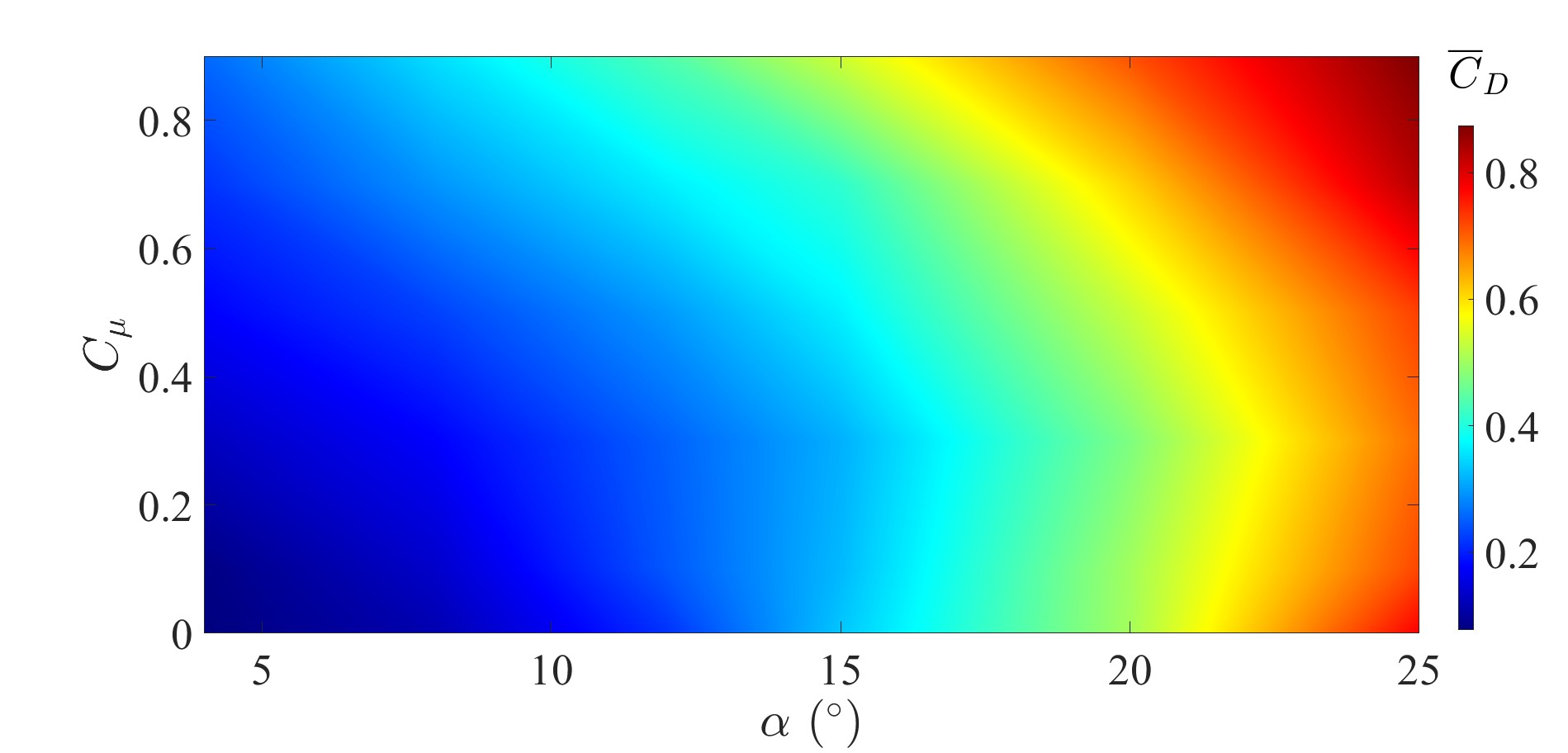}\label{fig:performancec}}
	\subfloat[]{\includegraphics[width=0.5 \textwidth]{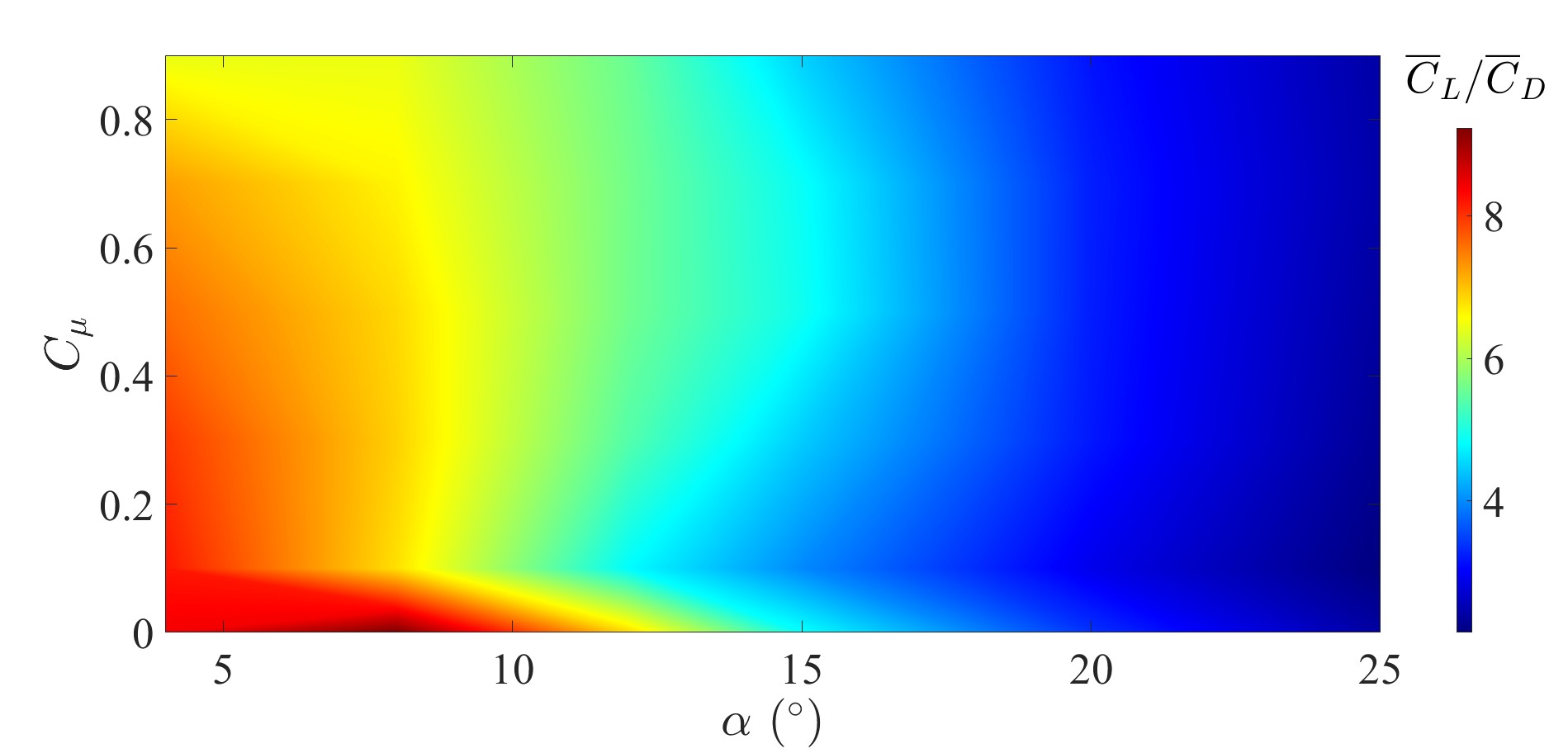}\label{fig:performanced}}
	\\
	\subfloat[]{\includegraphics[width=0.5 \textwidth]{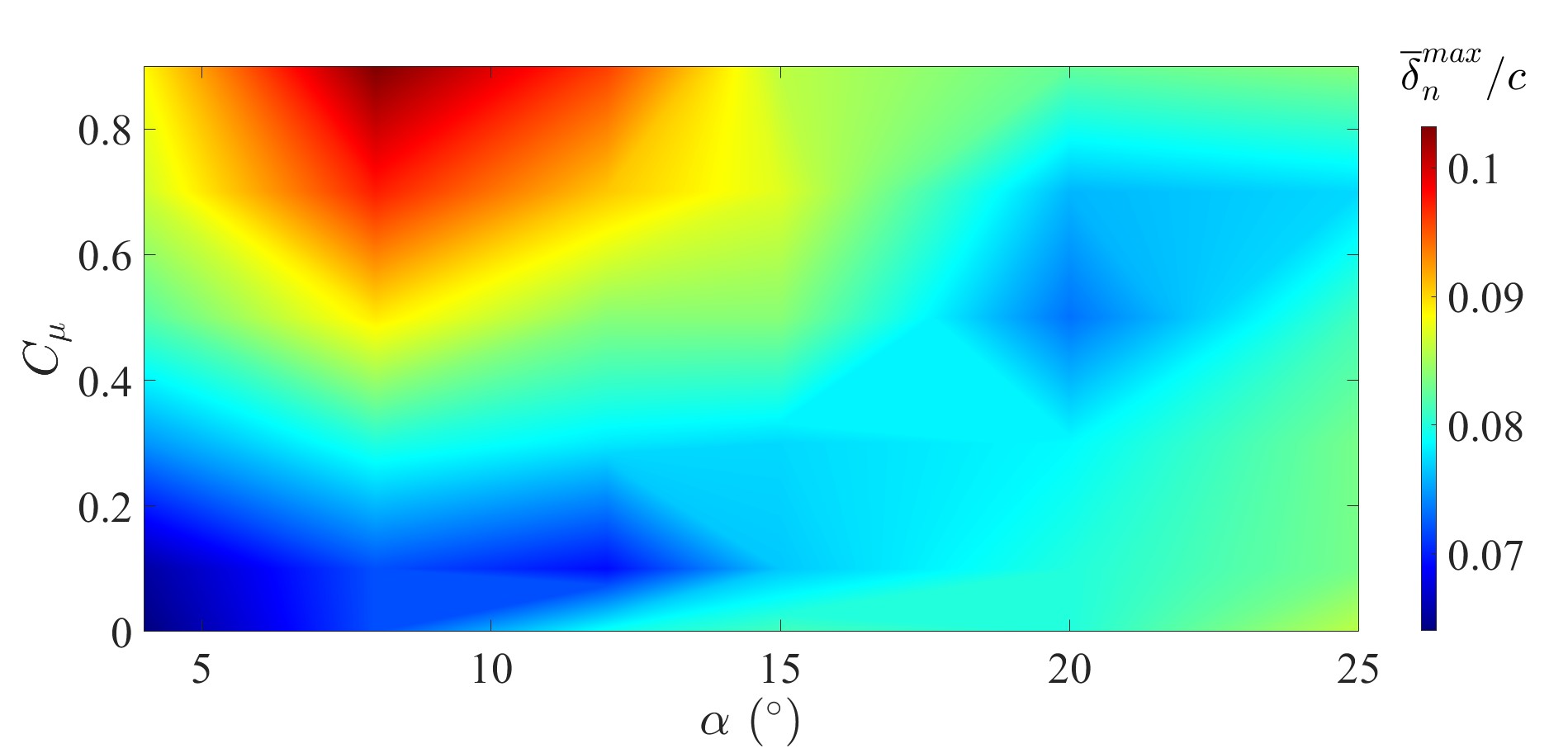}\label{fig:performancee}}
	\subfloat[]{\includegraphics[width=0.5 \textwidth]{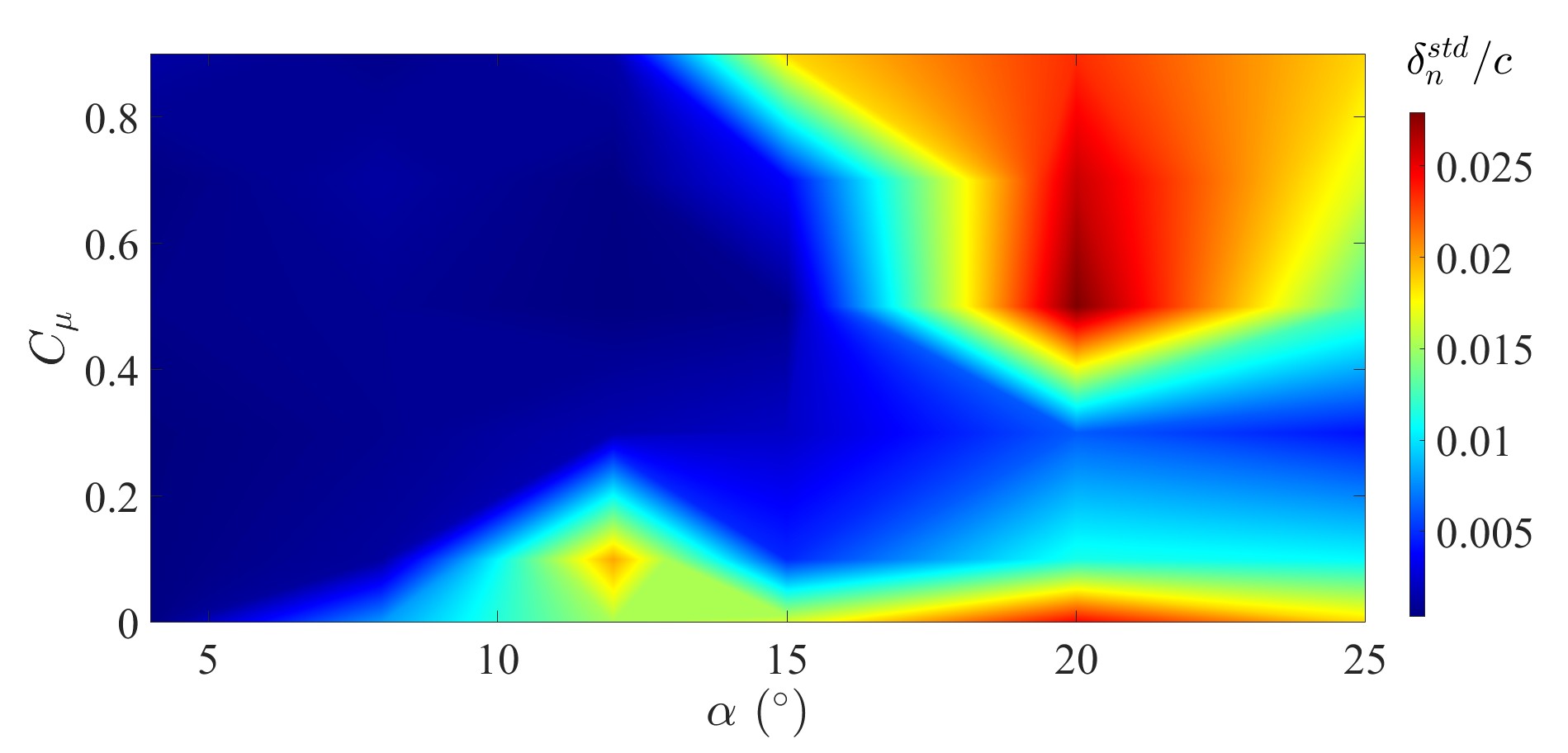}\label{fig:performancef}}
	\caption{\label{fig:performance}Comparison of (a) time-averaged lift coefficient, (b) standard deviation of lift coefficient, (c) time-averaged drag coefficient, (d) time-averaged lift-to-drag ratio, (e) maximum normal displacement and (f) standard deviation of membrane displacement in the parameter space of $\alpha-C_{\mu}$.}
\end{figure}

To examine the steady jet flow control effect, six groups of angles of attack and six sets of momentum coefficients are chosen to simulate the coupled fluid-membrane dynamics. The angle of attack ranges from $4^\circ$ up to a moderate value of $25^\circ$ to cover the cruise, take-off and landing state of flying vehicles. The momentum coefficient within [0, 0.9] is applied at the leading edge of the membrane to ensure that sufficient jet flows are injected into the wake. \refFig{fig:performance} presents the aerodynamic performance and the structural responses in the parameter space of $\alpha-C_{\mu}$.

\refFig{fig:performance} \subref{fig:performancea} plots the time-averaged lift coefficient in the examined parameter space. The lift coefficient keeps increasing at low angles of attack as a function of the momentum coefficient. When the angle of attack grows up and vortex shedding occurs, the lift performance degrades and then improves as the momentum coefficient increases. Optimal lift performance is achieved at $\alpha=12^\circ$ with the maximum momentum coefficient of $C_{\mu}=0.9$. Meanwhile, the lift fluctuation is suppressed once the jet flow control is applied at $\alpha < 15^\circ$ and under moderate momentum coefficients at relatively high angles of attack, as shown in \reffig{fig:performance} \subref{fig:performanceb}.

In \reffig{fig:performance} \subref{fig:performancec}, more drag penalties are achieved under the application of the active jet flow control. It can be seen from  \reffig{fig:performance} \subref{fig:performanced} that the current active jet flow control strategy is not beneficial to the lift-to-drag ratio performance. Optimal performance is achieved at $\alpha=8^\circ$ with no jet flows. Although the lift is improved under the control of jet flows, more drag synchronously increases, leading to poorer performance.

\refFigs{fig:performance} \subref{fig:performancee} and \subref{fig:performancef} present the maximum mean membrane displacement and the corresponding standard deviation value. The injected jet flows can help in increasing membrane deformation at $\alpha < 15^\circ$, thus improving mean lift coefficients. Similar distributions in the standard deviation contours of lift coefficient and membrane displacement can be observed. It can be inferred that the force fluctuations have a strong correlation with the structural vibrations.

\subsection{Effect of momentum coefficient} \label{sec:section4.2}
In this section, the coupled fluid-membrane dynamics at two representative angles of attack are compared systematically. The effects of the momentum coefficient on the instantaneous aerodynamic forces, the flow features and the membrane responses are analyzed.
\subsubsection{Membrane dynamics at low angle of attack}
\begin{figure}[H]
	\centering
	\subfloat[]{\includegraphics[width=0.5 \textwidth]{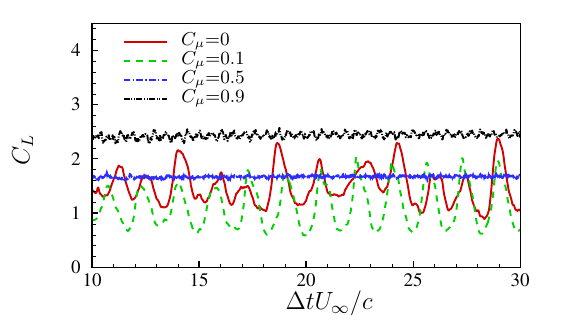}\label{fig:aoa12_forcea}}
	\subfloat[]{\includegraphics[width=0.5 \textwidth]{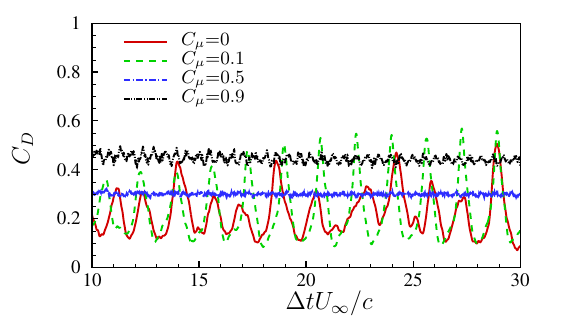}\label{fig:aoa12_forceb}}
	\caption{\label{fig:aoa12_force}Comparison of time-varying (a) lift coefficient and (b) drag coefficient at $\alpha$=$12^\circ$.}
\end{figure}

As discussed in \refse{sec:section4.1}, the flexible membrane with steady jet flow control exhibits different dynamical behaviors at various angles of attack. We first investigate the effect of the momentum coefficient at an angle of attack of $12^\circ$. \refFig{fig:aoa12_force} compares the time-varying lift and drag coefficients at this relatively low angle of attack. The instantaneous lift coefficient first decreases at $C_{\mu}$=0.1. Then, it is improved when more steady jet flows are injected into wake flows. The lift coefficient shows weaker fluctuations and higher frequencies at larger $C_{\mu}$ values. The instantaneous drag coefficient presents a similar variation trend as the momentum coefficient increases.

\begin{figure}[H]
	\centering
	\includegraphics[width=0.65 \textwidth]{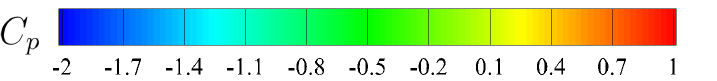}
	\\
	\subfloat[]{\includegraphics[width=1.0 \textwidth]{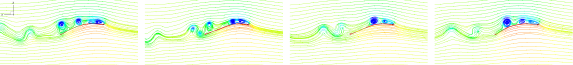}\label{fig:aoa12_streamlinea}}
	\\
	\subfloat[]{\includegraphics[width=1.0 \textwidth]{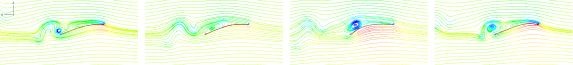}\label{fig:aoa12_streamlineb}}
	\\
	\subfloat[]{\includegraphics[width=1.0 \textwidth]{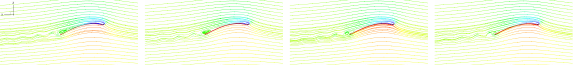}\label{fig:aoa12_streamlinec}}
	\\
	\subfloat[]{\includegraphics[width=1.0 \textwidth]{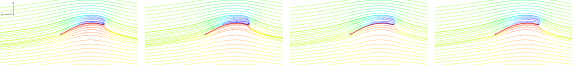}\label{fig:aoa12_streamlined}}
	\caption{\label{fig:aoa12_streamline}Comparison of instantaneous streamlines for $C_{\mu}$= (a) 0, (b) 0.1, (c) 0.5 and (d) 0.9 at $\alpha$=$12^\circ$.}
\end{figure}

\refFig{fig:aoa12_streamline} further presents a comparison of instantaneous streamlines at different momentum coefficients. A series of vortices are formed on the membrane's upper surface when no steady jet flow control is applied at the leading edge. These vortices with lower-pressure regions are blown away from the membrane's upper surface when high-momentum jet flows are injected. However, the injected jet flows are not strong enough to resist the adverse pressure gradient and suppress the vortex-shedding process at $C_{\mu}$=0.1. This weak steady jet leads to lower negative pressure on the upper surface and poorer lift performance. When the momentum coefficient increases up to 0.5, the vortex shedding process is almost suppressed and higher negative pressure is kept near the leading edge. Consequently, the flexible membrane exhibits better lift performance and weaker fluctuations in terms of aerodynamic forces and structural deformations. This trend becomes more obvious at a higher momentum coefficient in \reffig{fig:aoa12_streamline} \subref{fig:aoa12_streamlined}.

\begin{figure}[H]
	\centering
	\subfloat[]{\includegraphics[width=0.24 \textwidth]{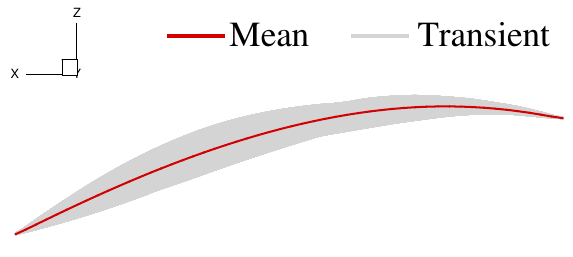}\label{fig:aoa12_envelopea}}
	\
	\subfloat[]{\includegraphics[width=0.24 \textwidth]{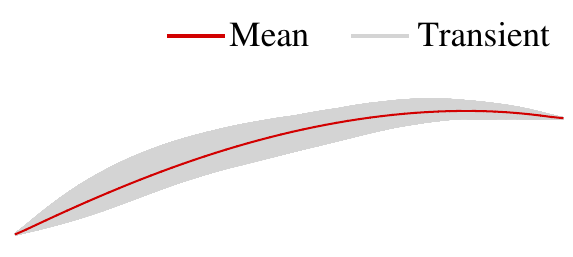}\label{fig:aoa12_envelopeb}}
	\
	\subfloat[]{\includegraphics[width=0.24 \textwidth]{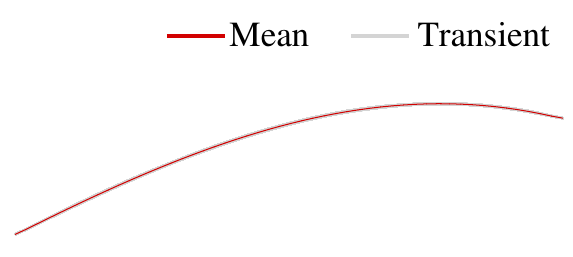}\label{fig:aoa12_envelopec}}
	\
	\subfloat[]{\includegraphics[width=0.24 \textwidth]{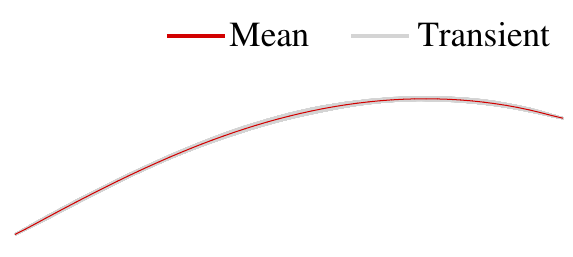}\label{fig:aoa12_enveloped}}
	\caption{\label{fig:aoa12_envelope}Comparison of instantaneous membrane vibration envelope for $C_{\mu}$= (a) 0, (b) 0.1, (c) 0.5 and (d) 0.9 at $\alpha$=$12^\circ$.}
\end{figure}

\begin{figure}[H]
	\centering
	\subfloat[]{\includegraphics[width=0.5 \textwidth]{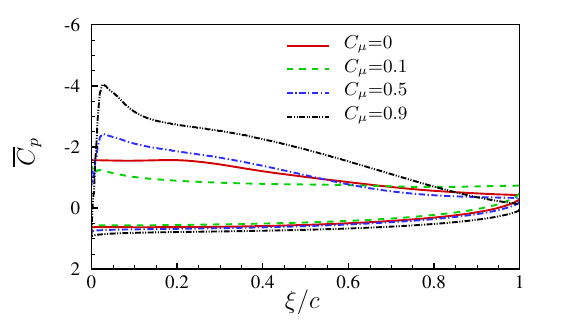}\label{fig:aoa12_disa}}
	\subfloat[]{\includegraphics[width=0.5 \textwidth]{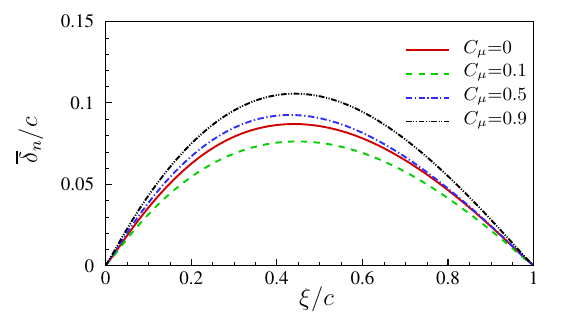}\label{fig:aoa12_disb}}
	\caption{\label{fig:aoa12_dis}Comparison of time-averaged (a) pressure coefficient and (b) normalized membrane displacement at $\alpha$=$12^\circ$.}
\end{figure}

A comparison of membrane vibration profiles at different time instants is shown in \reffig{fig:aoa12_envelope}. The flexible membrane exhibits obvious structural vibrations by synchronizing with the vortex shedding phenomenon under no or weak steady jet flow conditions. The flow-induced vibration of the flexible membrane is significantly suppressed when $C_{\mu}$ is greater than 0.3, as shown in \reffigs{fig:aoa12_envelope} \subref{fig:aoa12_envelopec} and \subref{fig:aoa12_enveloped}. \refFig{fig:aoa12_dis} presents the time-averaged pressure coefficient distributions and normalized membrane deformation along the membrane chord. It can be seen from \reffig{fig:aoa12_dis} \subref{fig:aoa12_disa} that the steady jet flows mainly alter the pressure distributions on the upper surface, but have less influence on those on the lower surface. The negative pressure near the leading edge is suppressed at a relatively small $C_{\mu}$ of 0.1. When the momentum coefficient increases up to 0.5, the negative pressure is enhanced, resulting in higher lift performance. The mean membrane deformation exhibits a reduced trend at $C_{\mu}$=0.1. Larger mean membrane deformations are achieved at higher momentum coefficients to balance additional aerodynamic loads induced by steady jet flows.

\begin{figure}[H]
	\centering
	\includegraphics[width=0.65 \textwidth]{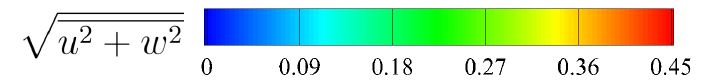}
	\\
	\subfloat[]{\includegraphics[width=0.24 \textwidth]{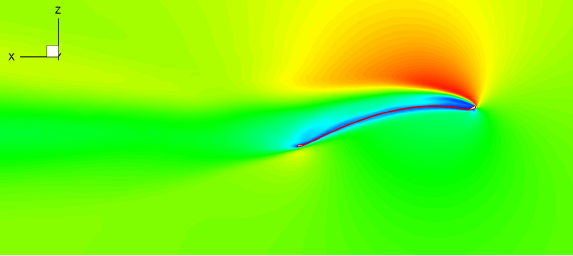}\label{fig:aoa12_flowa}}
	\
	\subfloat[]{\includegraphics[width=0.24 \textwidth]{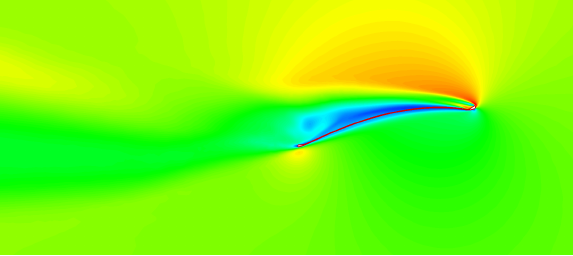}\label{fig:aoa12_flowb}}
	\
	\subfloat[]{\includegraphics[width=0.24 \textwidth]{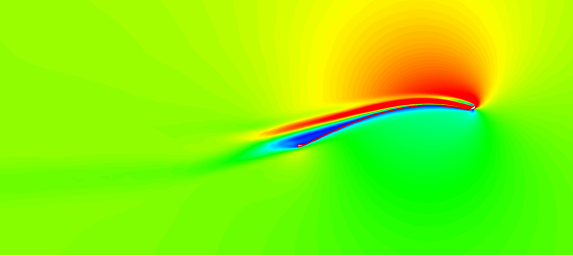}\label{fig:aoa12_flowc}}
	\
	\subfloat[]{\includegraphics[width=0.24 \textwidth]{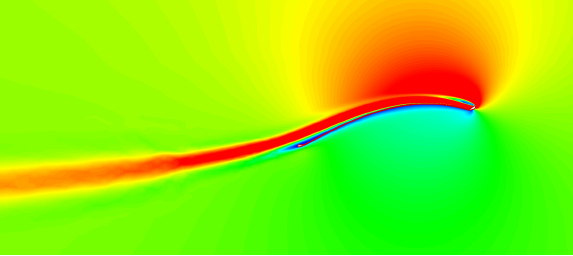}\label{fig:aoa12_flowd}}
	\\
	\includegraphics[width=0.65 \textwidth]{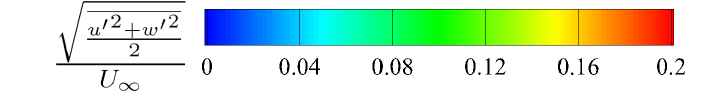}
	\\
	\subfloat[]{\includegraphics[width=0.24 \textwidth]{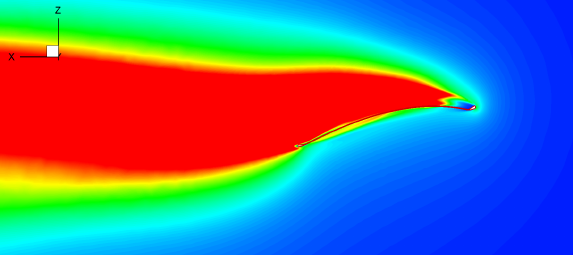}\label{fig:aoa12_flowe}}
	\
	\subfloat[]{\includegraphics[width=0.24 \textwidth]{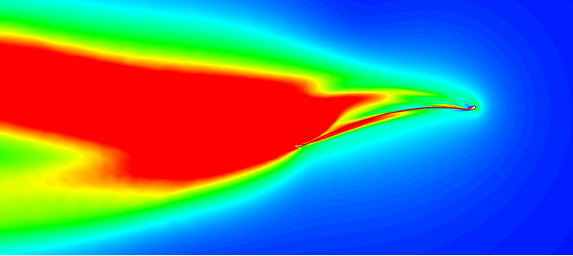}\label{fig:aoa12_flowf}}
	\
	\subfloat[]{\includegraphics[width=0.24 \textwidth]{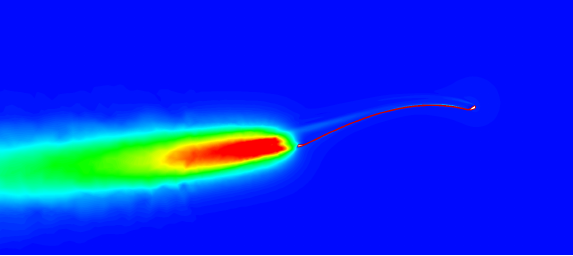}\label{fig:aoa12_flowg}}
	\
	\subfloat[]{\includegraphics[width=0.24 \textwidth]{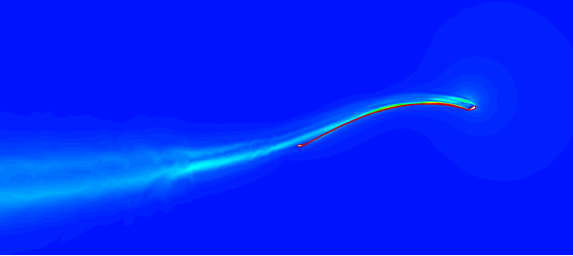}\label{fig:aoa12_flowh}}
	\caption{\label{fig:aoa12_flow}Comparison of (a,b,c,d) time-averaged velocity magnitude and (e,f,g,h) turbulence kinetic energy for $C_{\mu}$= (a,e) 0, (b,f) 0.1, (c,g) 0.5 and (d,h) 0.9 at $\alpha$=$12^\circ$.}
\end{figure}

The time-averaged velocity magnitude and the turbulence kinetic energy are extracted from the flow fields to further analyze the effect of steady jet flow, as shown in \reffig{fig:aoa12_flow}. The high momentum jet flows are elongated along the membrane chord at higher $C_{\mu}$, as observed from \reffigs{fig:aoa12_flow} (a-d). Due to the Coanda effect, the steady jet flows get attached to the cambered membrane surface. By comparing the turbulence kinetic energy in \reffigs{fig:aoa12_flow} (e-h), the flow fluctuations in the wake behind the flexible membrane are significantly reduced. It can be inferred that the high-momentum jet flows are transported into the boundary layer on the upper surface, which carry enough energy to resist the formation of shedding vortices. Finally, the flow fluctuations and the flow-induced vibrations are suppressed.

\subsubsection{Membrane dynamics at high angle of attack}
\begin{figure}[H]
	\centering
	\subfloat[]{\includegraphics[width=0.5 \textwidth]{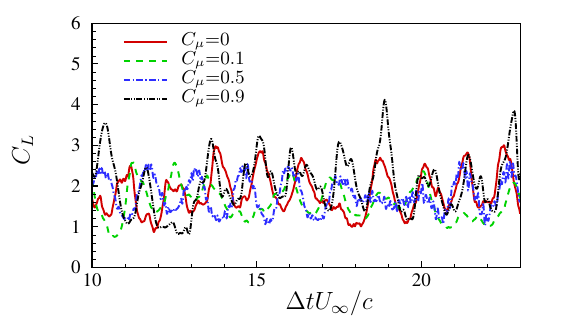}\label{fig:aoa25_forcea}}
	\subfloat[]{\includegraphics[width=0.5 \textwidth]{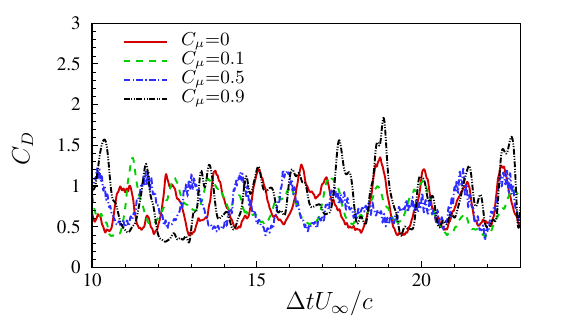}\label{fig:aoa25_forceb}}
	\caption{\label{fig:aoa25_force}Comparison of time-varying (a) lift coefficient and (b) drag coefficient at $\alpha$=$25^\circ$.}
\end{figure}

We next investigate the coupled fluid-membrane dynamics at a relatively high angle of attack of $25^\circ$. Bats and flying vehicles typically perform landing, take-off and maneuvers operations at this high angle of attack. Large-scale vortices are generated behind the wing and negative pressure is dropped, which limits the flight efficiency at this representative angle of attack. Herein, the aerodynamic performance and the structural responses with steady jet flow control at $\alpha$=$25^\circ$ are examined. \refFig{fig:aoa25_force} presents the time-varying lift and drag coefficients at four selected momentum coefficients. Both lift and drag forces first show an overall downward trend, and then increase slightly as the momentum coefficient increases. The fluctuation amplitude becomes smaller at $C_{\mu}$=0.1. When the momentum coefficient keeps increasing, the fluctuation of the aerodynamic loads gets more severe.

\begin{figure}[H]
	\centering
	\includegraphics[width=0.65 \textwidth]{./results/effect_cu/aoa12/cp_leng.pdf}
	\\
	\subfloat[]{\includegraphics[width=1.0 \textwidth]{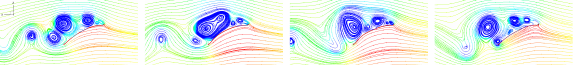}\label{fig:aoa25_streamlinea}}
	\\
	\subfloat[]{\includegraphics[width=1.0 \textwidth]{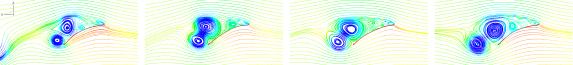}\label{fig:aoa25_streamlineb}}
	\\
	\subfloat[]{\includegraphics[width=1.0 \textwidth]{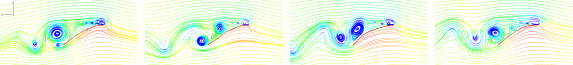}\label{fig:aoa25_streamlinec}}
	\\
	\subfloat[]{\includegraphics[width=1.0 \textwidth]{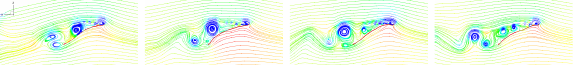}\label{fig:aoa25_streamlined}}
	\caption{\label{fig:aoa25_streamline}Comparison of instantaneous streamlines for $C_{\mu}$= (a) 0, (b) 0.1, (c) 0.5 and (d) 0.9 at $\alpha$=$25^\circ$.}
\end{figure}

\begin{figure}[H]
	\centering
	\subfloat[]{\includegraphics[width=0.24 \textwidth]{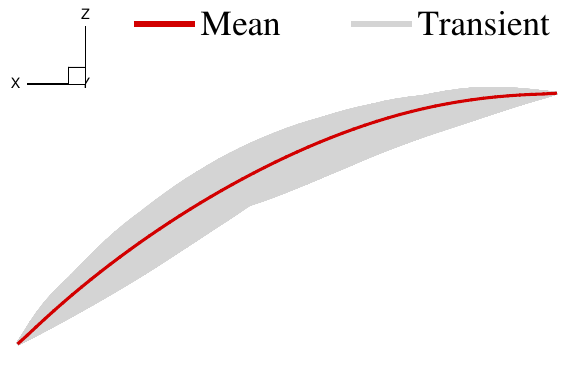}\label{fig:aoa25_envelopea}}
	\
	\subfloat[]{\includegraphics[width=0.24 \textwidth]{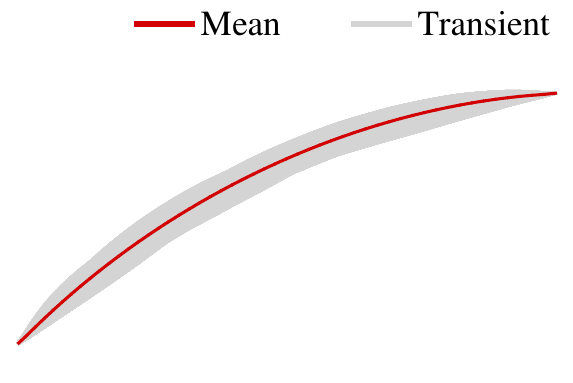}\label{fig:aoa25_envelopeb}}
	\
	\subfloat[]{\includegraphics[width=0.24 \textwidth]{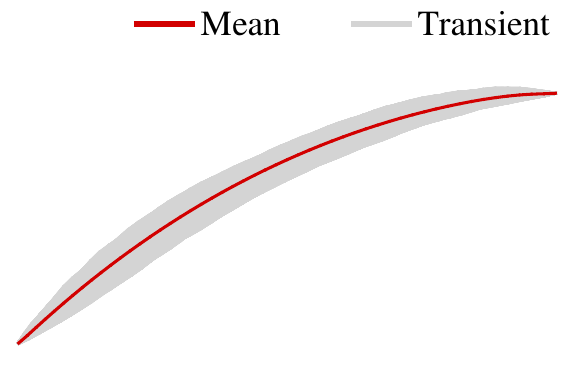}\label{fig:aoa25_envelopec}}
	\
	\subfloat[]{\includegraphics[width=0.24 \textwidth]{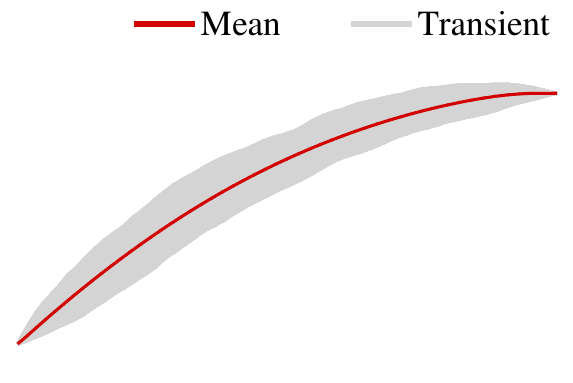}\label{fig:aoa25_enveloped}}
	\caption{\label{fig:aoa25_envelope}Comparison of instantaneous membrane vibration envelope for $C_{\mu}$= (a) 0, (b) 0.1, (c) 0.5 and (d) 0.9 at $\alpha$=$25^\circ$.}
\end{figure}

\refFig{fig:aoa25_streamline} compares the instantaneous streamlines at four momentum coefficients. Large-scale vortices are formed on the leeward side of the flexible membrane when there is no jet flow injection. The separated flows are suppressed as the high-energy jet flows are transported and mixed with the unsteady flows near the leading edge. As shown in \reffig{fig:aoa25_streamline} \subref{fig:aoa25_streamlineb}, the negative pressure regions are reduced at $C_{\mu}$=0.1, leading to overall smaller lift performance. Some small-scale vortices and streamline oscillations are noticed along the jet flow path near the leading edge at a higher momentum coefficient. These phenomena are mainly caused by the Rayleigh-Taylor instability of the high-speed jet flows. The shedding vortices are not suppressed by the steady jet flows at $\alpha$=$25^\circ$. The flow-induced vibration of the flexible membrane is still triggered and becomes more severe at a higher momentum coefficient, which is compared in \reffig{fig:aoa25_envelope}.

\begin{figure}[H]
	\centering
	\subfloat[]{\includegraphics[width=0.5 \textwidth]{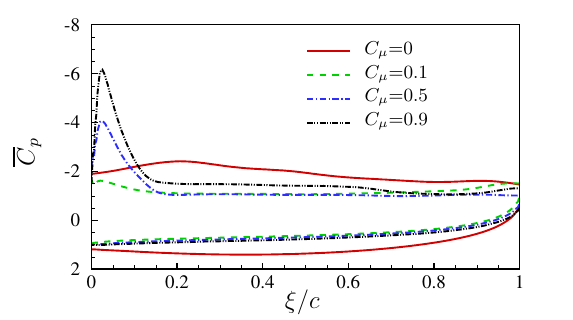}\label{fig:aoa25_disa}}
	\subfloat[]{\includegraphics[width=0.5 \textwidth]{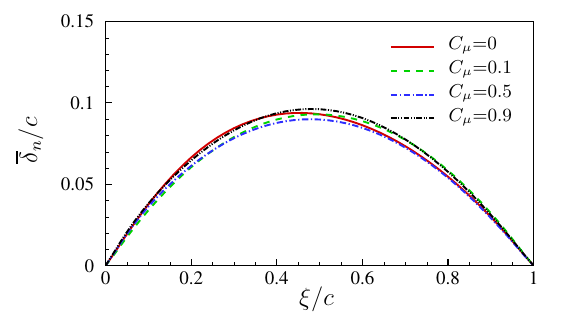}\label{fig:aoa25_disb}}
	\caption{\label{fig:aoa25_dis}Comparison of time-averaged (a) pressure coefficient and (b) normalized membrane displacement at $\alpha$=$25^\circ$.}
\end{figure}

We further compare the time-averaged pressure coefficient distributions and the normalized membrane deformation along the membrane chord in \reffig{fig:aoa25_dis}. As consistent with the results shown in \reffig{fig:aoa25_streamline}, the negative pressure on the leeward side of the membrane is weakened when steady jet flow with $C_{\mu}$=0.1 is applied at the leading edge. The negative pressure region is enhanced near the leading edge as the momentum coefficient increases up to 0.5. The steady jet flow control shows little effect on the membrane deformation, as observed from \reffig{fig:aoa25_dis} \subref{fig:aoa25_disb}.

\begin{figure}[H]
	\centering
	\includegraphics[width=0.65 \textwidth]{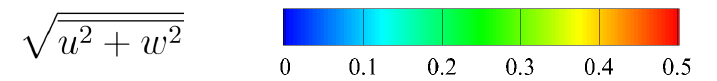}
	\\
	\subfloat[]{\includegraphics[width=0.24 \textwidth]{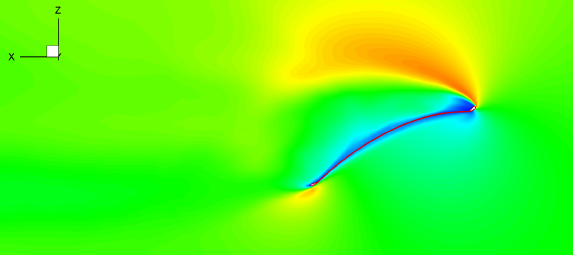}\label{fig:aoa25_flowa}}
	\
	\subfloat[]{\includegraphics[width=0.24 \textwidth]{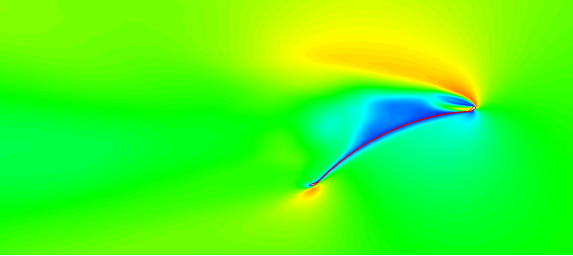}\label{fig:aoa25_flowb}}
	\
	\subfloat[]{\includegraphics[width=0.24 \textwidth]{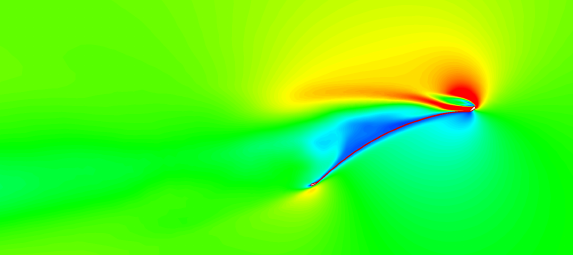}\label{fig:aoa25_flowc}}
	\
	\subfloat[]{\includegraphics[width=0.24 \textwidth]{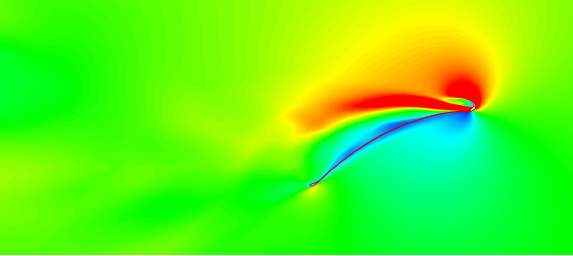}\label{fig:aoa25_flowd}}
	\\
	\includegraphics[width=0.65 \textwidth]{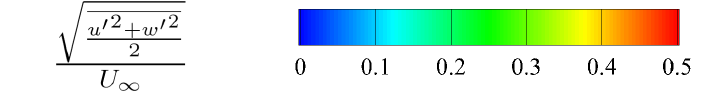}
	\\
	\subfloat[]{\includegraphics[width=0.24 \textwidth]{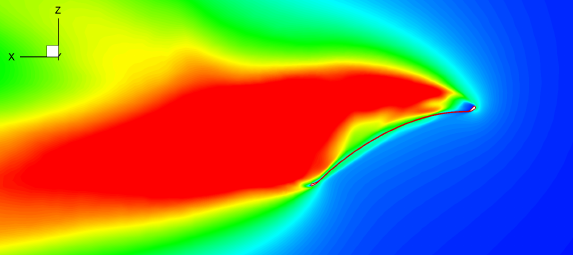}\label{fig:aoa25_flowe}}
	\
	\subfloat[]{\includegraphics[width=0.24 \textwidth]{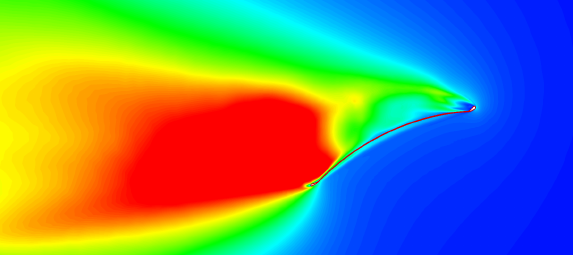}\label{fig:aoa25_flowf}}
	\
	\subfloat[]{\includegraphics[width=0.24 \textwidth]{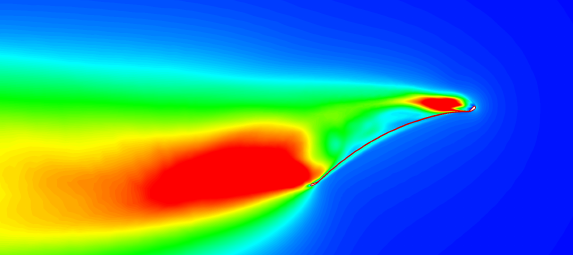}\label{fig:aoa25_flowg}}
	\
	\subfloat[]{\includegraphics[width=0.24 \textwidth]{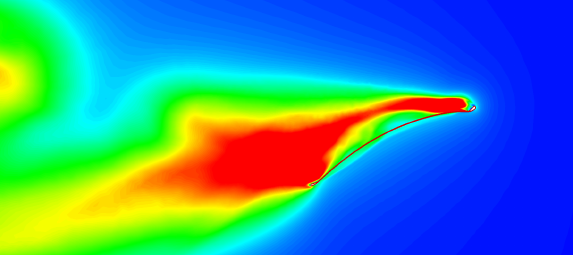}\label{fig:aoa25_flowh}}
	\caption{\label{fig:aoa25_flow}Comparison of (a,b,c,d) time-averaged velocity magnitude and (e,f,g,h) turbulence kinetic energy for $C_{\mu}$= (a,e) 0, (b,f) 0.1, (c,g) 0.5 and (d,h) 0.9 at $\alpha$=$25^\circ$.}
\end{figure}

A comparison of the time-averaged velocity magnitude at four momentum coefficients is presented in \reffig{fig:aoa25_flow} (a-d). The steady jet flows with high momentum are not attached to the upper surface of the flexible membrane. It shows significant differences from the jet flow paths close to the membrane surface at the low angle of attack of $12^\circ$. The Coanda effect does not dominate the jet flow path at high angles of attack. The main reason is that the negative pressure in the wake flows recovers to ambient pressure quickly under the influence of the steady jet flow at high angles of attack. Thus, the jet flow path is enforced to keep away from the membrane surface by the recovered pressure in the vicinity of the leeward surface. \refFigs{fig:aoa25_flow} (e-h) present the turbulence kinetic energy at four momentum coefficients. High flow fluctuation regions are attached to the membrane's upper surface even with steady jet flow control. Since the high-momentum jet flows are not transported into the low-momentum boundary layer flows, the wake flows are still perturbed by the shedding vortices. The flexible membrane is forced to vibrate through the fluid-structure coupling effect, which is consistent with the observations in \reffig{fig:aoa25_envelope}.

\subsection{Effect of jet flow control on aerodynamic loads} \label{sec:section4.3}
The investigations of the membrane dynamics in \refse{sec:section4.2} confirm that the steady jet flows possess the ability to adjust the aerodynamic performance. In this section, the effect of jet flow control on aerodynamic loads is examined in detail. Two nondimensional parameters, namely lift-gain factor (LGF) and drag-penalty factor (DPF) are defined to evaluate the variation of the aerodynamic loads under jet flow control. In \refeq{LGF}, the gain or penalty factors of the aerodynamic forces are calculated as the ratio between the force coefficient variation relative to a reference at $C_{\mu}$=0 and the momentum coefficient, which are given as

\begin{equation}
	C_{LGF}=\frac{\overline{C}_L-\overline{C}_L^{ref}}{C_{\mu}}, \quad \quad C_{DPF}=\frac{\overline{C}_D-\overline{C}_D^{ref}}{C_{\mu}}
	\label{LGF} 
\end{equation}

\begin{figure}[H]
	\centering
	\subfloat[]{\includegraphics[width=0.5 \textwidth]{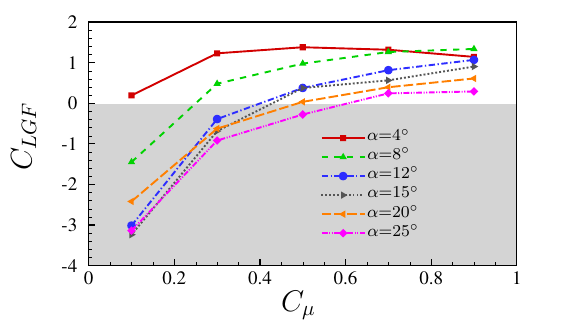}\label{fig:LGFa}}
	\subfloat[]{\includegraphics[width=0.5 \textwidth]{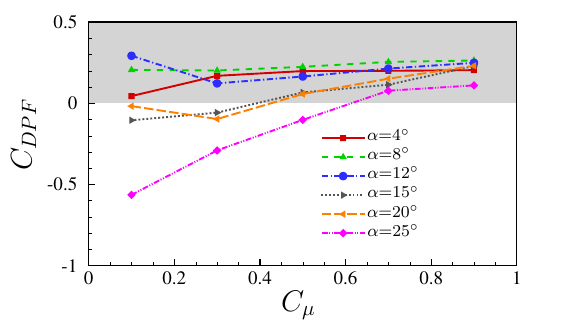}\label{fig:LGFb}}
	\caption{\label{fig:LGF} Comparison of (a) lift-gain factor and (b) drag-penalty factor as a function of momentum coefficient.}
\end{figure}

\refFig{fig:LGF} presents the lift-gain factor and the drag-penalty factor as a function of the momentum coefficient. The white part in \reffig{fig:LGF} indicates the lift improvement or the drag reduction compared to the aerodynamic loads at $C_{\mu}$=0. It can be seen from \reffig{fig:LGF} that both factors exhibit an overall upward trend as a function of the momentum coefficient. This upward trend indicates that the high-momentum jet flows can adjust the aerodynamic loads by mixing with the wake flows behind the flexible membrane. The steady jet flow control can help in improving the lift performance at $\alpha$=$4^\circ$ once it is applied at the leading edge. When the angle of attack increases, higher momentum coefficients are required to achieve the lift gain. On the contrary, steady jet flows with smaller momentum coefficients are beneficial to drag reduction at larger angles of attack.

\begin{figure}[H]
	\centering
	\subfloat[]{\includegraphics[width=0.5 \textwidth]{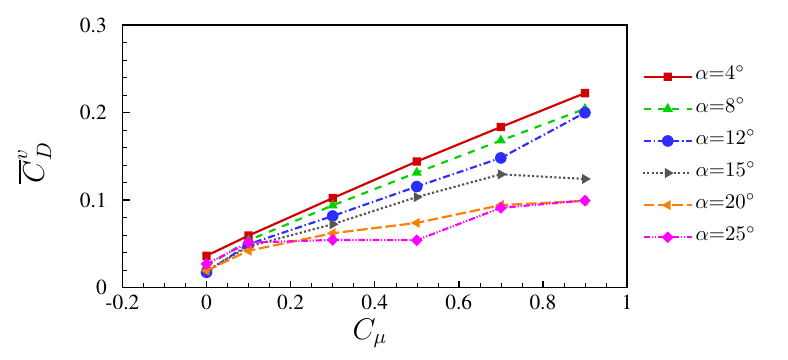}\label{fig:draga}}
	\subfloat[]{\includegraphics[width=0.5 \textwidth]{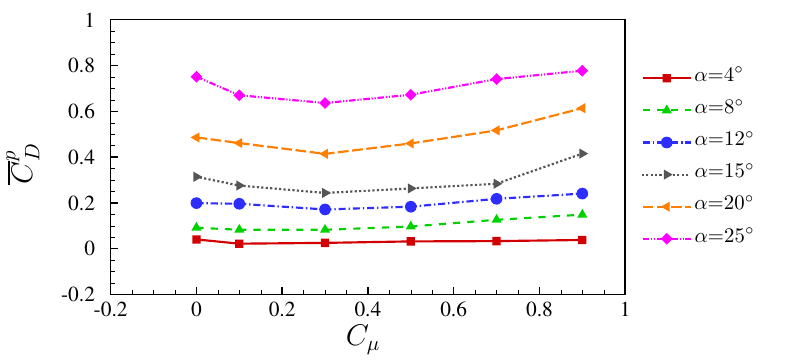}\label{fig:dragb}}
	\\
	\subfloat[]{\includegraphics[width=0.5 \textwidth]{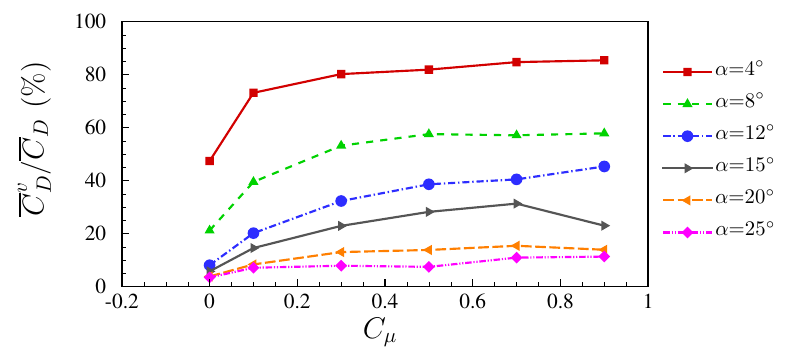}\label{fig:dragc}}
	\subfloat[]{\includegraphics[width=0.5 \textwidth]{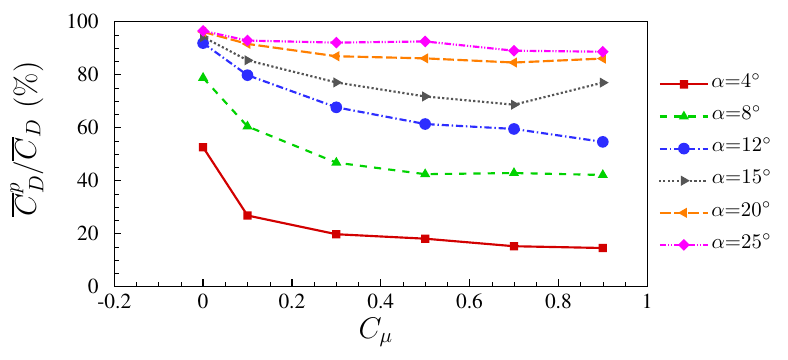}\label{fig:dragd}}
	\caption{\label{fig:drag}Decomposition of time-averaged drag coefficient into (a) viscous drag coefficient and (b) pressure drag coefficient. Ratio of (c) viscous drag coefficient and (d) pressure drag coefficient to total drag coefficient.}
\end{figure}

The total drag coefficient $C_D$ can be decomposed into the viscous drag coefficient $C_D^v$ and the pressure drag coefficient $C_D^p$ based on the sources of drag generation. In \reffig{fig:drag}, the effects of jet flow control on the variation of different drag components and their percentages of the total drag are further examined. As shown in \reffig{fig:drag} \subref{fig:draga}, the steady jet flows increase the contribution of the viscous force to the drag generation. The pressure drag coefficient reduces first and then grows as the momentum coefficient increases. In \reffigs{fig:drag} (c-d), the contribution of the viscous drag to the total drag exhibits an overall upward trend, while the ratio of the pressure drag decreases as a function of the momentum coefficient. The viscous drag component is dominated in the total drag generation at low angles of attack. The pressure drag part gradually plays a major role when the angle of attack exceeds $12^\circ$. The high-momentum jet flows are transported into the boundary layer on the leeward side of the flexible membrane and greatly increase the viscous forces. Due to the mix of the wake flows and the jet flows, the pressure distributions in the vicinity of the membrane are changed by adjusting the vortex patterns.

\subsection{Mechanism of aerodynamic performance improvement and vibration suppression} \label{sec:section4.4}
Based on the results presented in previous sections, the steady jet flow control shows its capability to adjust the aerodynamic performance and suppress the flow-induced vibration. Some underlying mechanisms must exist in the jet flow-controlled system to govern the coupled fluid-membrane dynamics. The jet flows with high momentum are transported into the wake flows to adjust the vortex-shedding patterns. The pressure distribution and the structural response are changed accordingly through the fluid-structure coupling effect. Therefore, the vortex patterns and the high-momentum jet flow transport process can be used as an entry point to explore the physical mechanism. In this section, the POD method presented in \refse{sec:section2.3} is used to extract the dominant vortex patterns of interest from the coupled fluid-membrane system. The flow transport process is analyzed from the viewpoint of Lagrangian Coherent Structures. The relationship between the aerodynamic loads and the structural deformations is revealed by two scaling relations.

\begin{figure}[H]
	\centering
	\includegraphics[width=0.65 \textwidth]{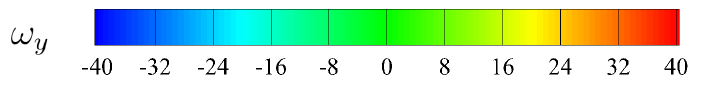}
	\\
	\subfloat[]{\includegraphics[width=0.24 \textwidth]{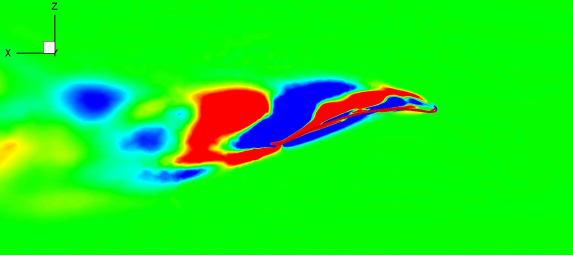}\label{fig:PODa}}
	\
	\subfloat[]{\includegraphics[width=0.24 \textwidth]{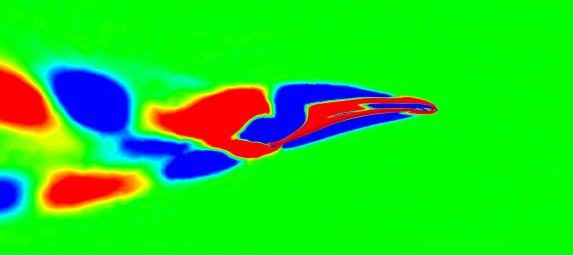}\label{fig:PODb}}
	\
	\subfloat[]{\includegraphics[width=0.24 \textwidth]{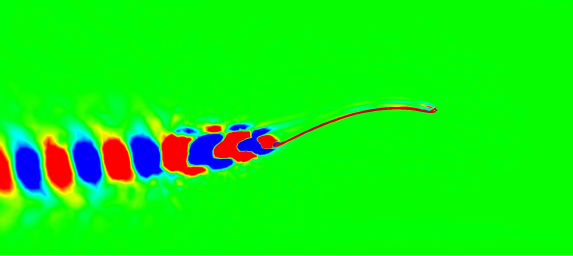}\label{fig:PODc}}
	\
	\subfloat[]{\includegraphics[width=0.24 \textwidth]{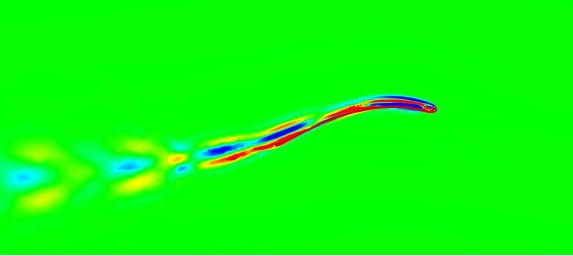}\label{fig:PODd}}
	\\
	\subfloat[]{\includegraphics[width=0.24 \textwidth]{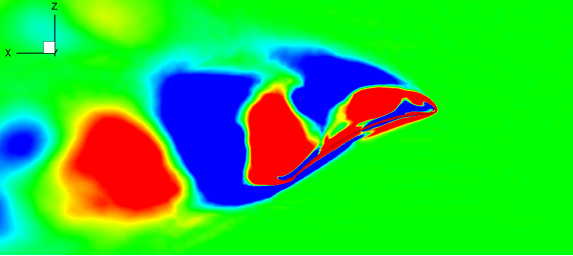}\label{fig:PODe}}
	\
	\subfloat[]{\includegraphics[width=0.24 \textwidth]{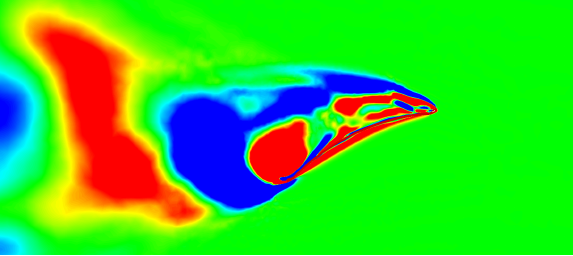}\label{fig:PODf}}
	\
	\subfloat[]{\includegraphics[width=0.24 \textwidth]{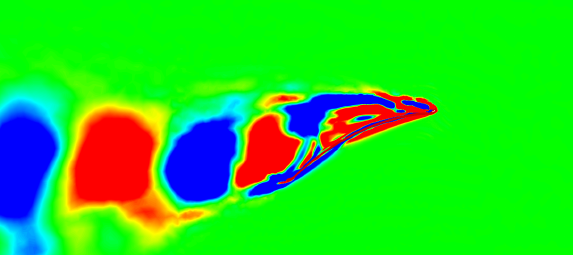}\label{fig:PODg}}
	\
	\subfloat[]{\includegraphics[width=0.24 \textwidth]{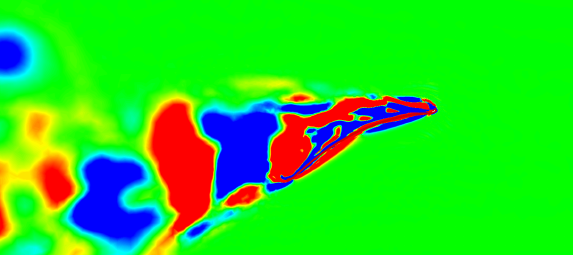}\label{fig:PODh}}
	\caption{\label{fig:POD}Comparison of dominant POD modes for $Y$-vorticity fields at $\alpha$= (a,b,c,d) $12^\circ$ and (e,f,g,h) $25^\circ$ for $C_{\mu}$= (a,e) 0, (b,f) 0.1, (c,g) 0.5 and (d,h) 0.9.}
\end{figure}

In \reffig{fig:POD}, the dominant POD modes extracted from the $Y$-vorticity fields are presented at two representative angles of attack for different momentum coefficients. The vortices are formed near the leading edge and shed into the wake in a Strouhal number around 0.52 at $\alpha$=$12^\circ$ for low momentum coefficients. As $C_{\mu}$ increases to 0.5, the vortices are generated near the trailing edge with a Strouhal number nearly three times of that at $C_{\mu}$=0. The vortex shedding phenomenon is almost eliminated by a high momentum coefficient of $C_{\mu}$=0.9. The large-scale vortices are perturbed into smaller-size vortices when the steady jet flows are applied at a relatively larger angle of attack of $\alpha$=$25^\circ$. The vortex pattern is not affected significantly by the momentum coefficient higher than 0.5.

\begin{figure}[H]
	\centering
	\subfloat[]{\includegraphics[width=0.45 \textwidth]{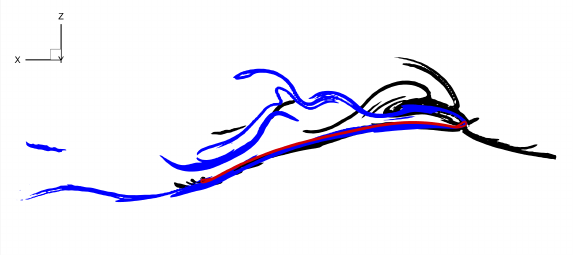}\label{fig:LCS12a}}
	\
	\subfloat[]{\includegraphics[width=0.45 \textwidth]{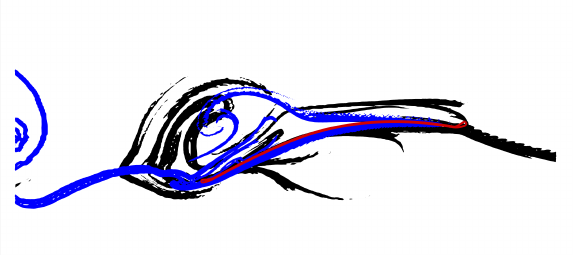}\label{fig:LCS12b}}
	\\
	\subfloat[]{\includegraphics[width=0.45 \textwidth]{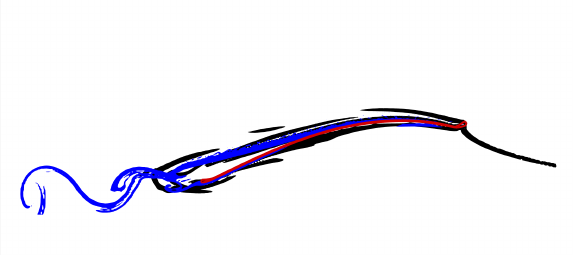}\label{fig:LCS12c}}
	\
	\subfloat[]{\includegraphics[width=0.45 \textwidth]{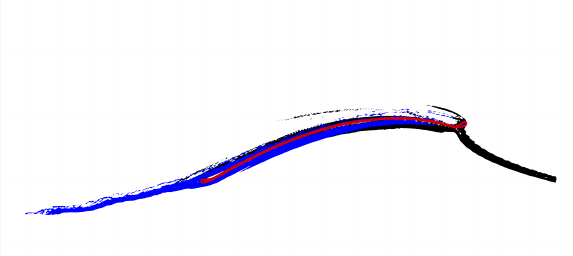}\label{fig:LCS12d}}
	\caption{\label{fig:LCS12}Comparison of Lagrangian Coherent Structures at $\alpha$=$12^\circ$for $C_{\mu}$= (a) 0, (b) 0.1, (c) 0.5 and (d) 0.9. The blue structures represent attracting LCSs and the black structures denote  repelling LCSs. The red part is the flexible membrane profile.}
\end{figure}

\refFig{fig:LCS12} presents the LCSs at $\alpha$=$12^\circ$ for four selected momentum coefficients. The attracting LCSs in blue color and the repelling LCSs in black color are plotted together to reflect the fluid transport process. The instantaneous flexible membrane profile at the extracted time instant is added to distinguish the structural boundary. As shown in \reffig{fig:LCS12} \subref{fig:LCS12a}, the attracting LCSs and the repelling LCSs have high coincidence at the leading edge. The material exchange between the freestream flows and the low-momentum flows in the separation region near the leading edge is very slight. Vortices are generated in this region and gradually move downward. Tangles between the repelling LCSs and the attracting LCSs are observed downstream of the leading edge separation. This phenomenon indicates that the fluid in the freestream flows is transported into the separation regions, resulting in periodic vortex shedding. 

When high-momentum jet flows are applied at the leading edge shown in \reffig{fig:LCS12} \subref{fig:LCS12b}, jet flows are transported through a tunnel between the attracting LCSs and the repelling LCSs near the leading edge to the separation region and eliminate the leading edge vortices presented in \reffig{fig:LCS12} \subref{fig:LCS12a}. Thus, the momentum of the jet flows is greatly consumed by these low-momentum separated flows. Due to the strong adverse pressure gradient, the jet flows no longer have excess momentum to be transported to the separation region near the trailing edge. Two types of LCSs are intersected in the rear part of the membrane to form the vortex shedding from the trailing edge. As the momentum coefficient increases to $C_{\mu}$=0.5, more momentum is carried by jet flows to overcome the strong adverse pressure gradient caused by the separation region formed in the middle of the membrane displayed in \reffig{fig:LCS12} \subref{fig:LCS12b}. Consequently, the fluid transport process between the jet flows and the separation region is enhanced and the large vortex in the middle of the membrane is blown away. Small-scale vortices are formed at the trailing edge and shed into the wake alternatively. When the momentum coefficient is high enough, the jet flows are attached to the membrane surface due to the Coanda effect as plotted in \reffig{fig:LCS12} \subref{fig:LCS12d}. Both types of LCSs overlap each other and almost no vortex is generated.

\begin{figure}[H]
	\centering
	\subfloat[]{\includegraphics[width=0.45 \textwidth]{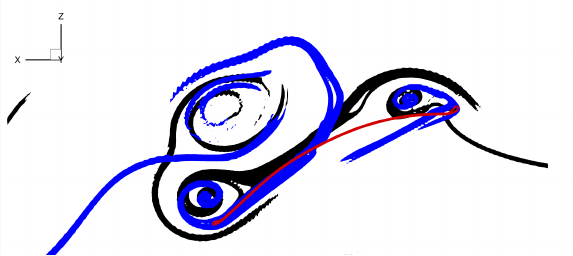}\label{fig:LCS25a}}
	\
	\subfloat[]{\includegraphics[width=0.45 \textwidth]{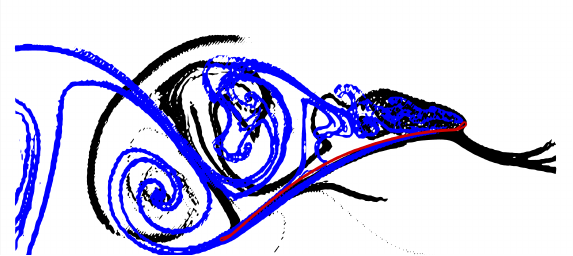}\label{fig:LCS25b}}
	\\
	\subfloat[]{\includegraphics[width=0.45 \textwidth]{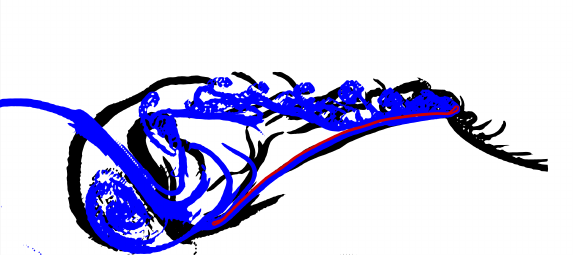}\label{fig:LCS25c}}
	\
	\subfloat[]{\includegraphics[width=0.45 \textwidth]{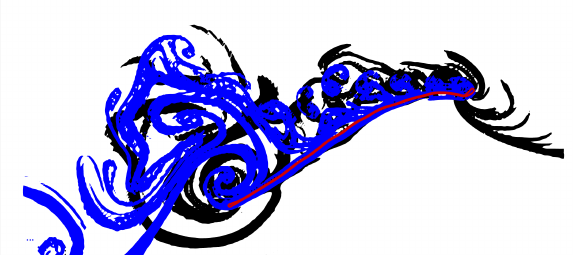}\label{fig:LCS25d}}
	\caption{\label{fig:LCS25}Comparison of Lagrangian Coherent Structures at $\alpha$=$25^\circ$for $C_{\mu}$= (a) 0, (b) 0.1, (c) 0.5 and (d) 0.9. The blue structures represent attracting LCSs and the black structures denote  repelling LCSs. The red part is the flexible membrane profile.}
\end{figure}

As the angle of attack increases to $\alpha$=$25^\circ$ with massive separated flows, the fluid transport process becomes more complex as a function of the momentum coefficient. In \reffig{fig:LCS25} \subref{fig:LCS25a}, the freestream flows are transported into the separation region near the leading edge and vortices are generated. The materials inside the separation area in the middle of the membrane are transported to the main flows, leading to vortex shedding. When the jet flow control is applied at the leading edge, the attracting LCSs and the repelling LCSs are tangled along the freestream direction. The tangled area is elongated downward as the momentum coefficient increases. Material exchange occurs between the main flows and the separated flows, leading to vortex shedding near the trailing edge. At high angles of attack, the strong adverse pressure gradient enforces the jet flows away from the membrane surface, which weakens the fluid transport to the boundary layer. As a result, the flow separation is not suppressed by the jet flow control at high angles of attack. The flow features show similarities to those behind bluff bodies (like circular cylinders). The active jet flow actuators placed at the leading edge are not the optimal choice to achieve flow separation suppression. Conversely, the actuators placed near the middle of the membrane are beneficial to aerodynamic performance improvement.

\begin{figure}[H]
	\centering
	\subfloat[]{\includegraphics[width=0.95 \textwidth]{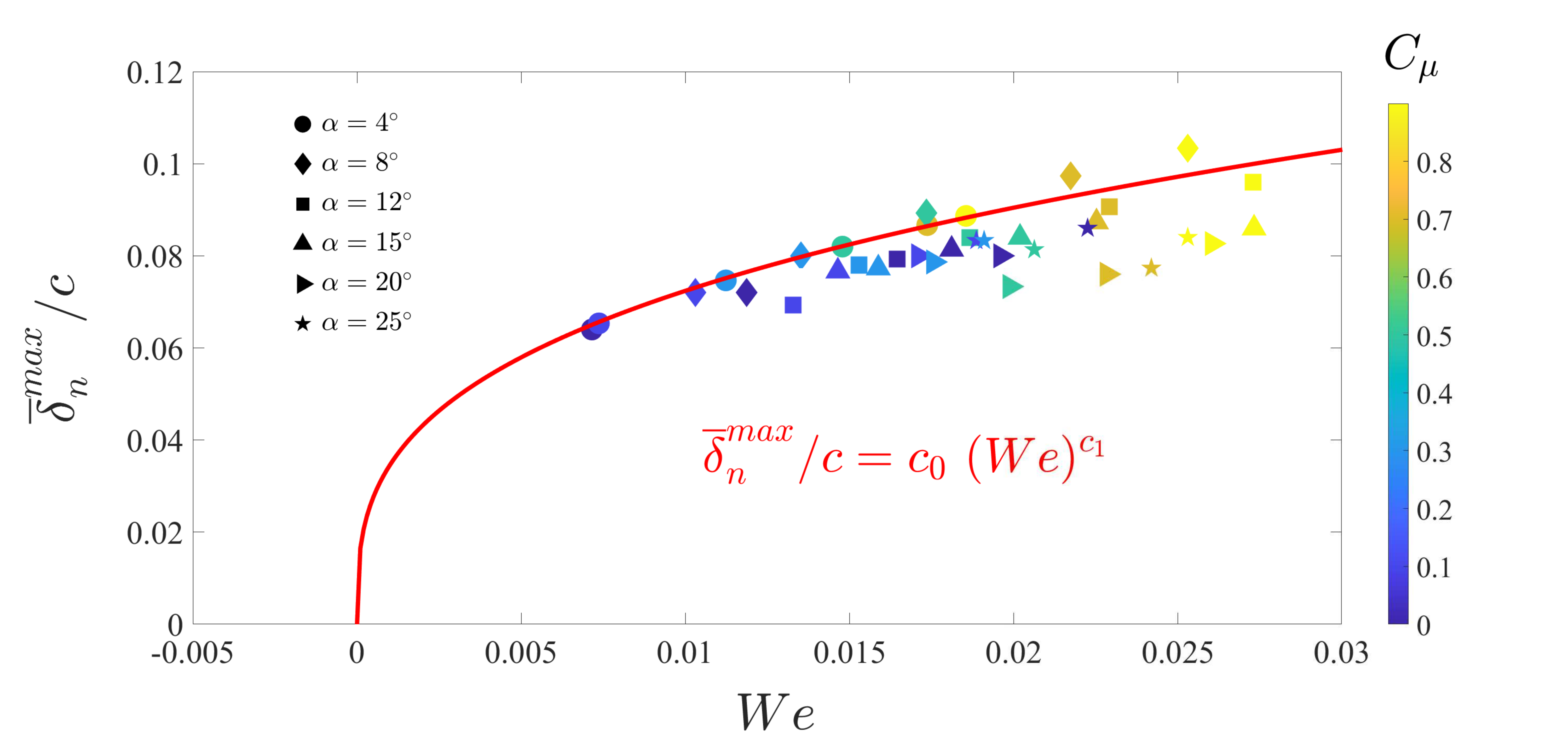}\label{fig:dis_wea}}
	\\
	\subfloat[]{\includegraphics[width=0.95 \textwidth]{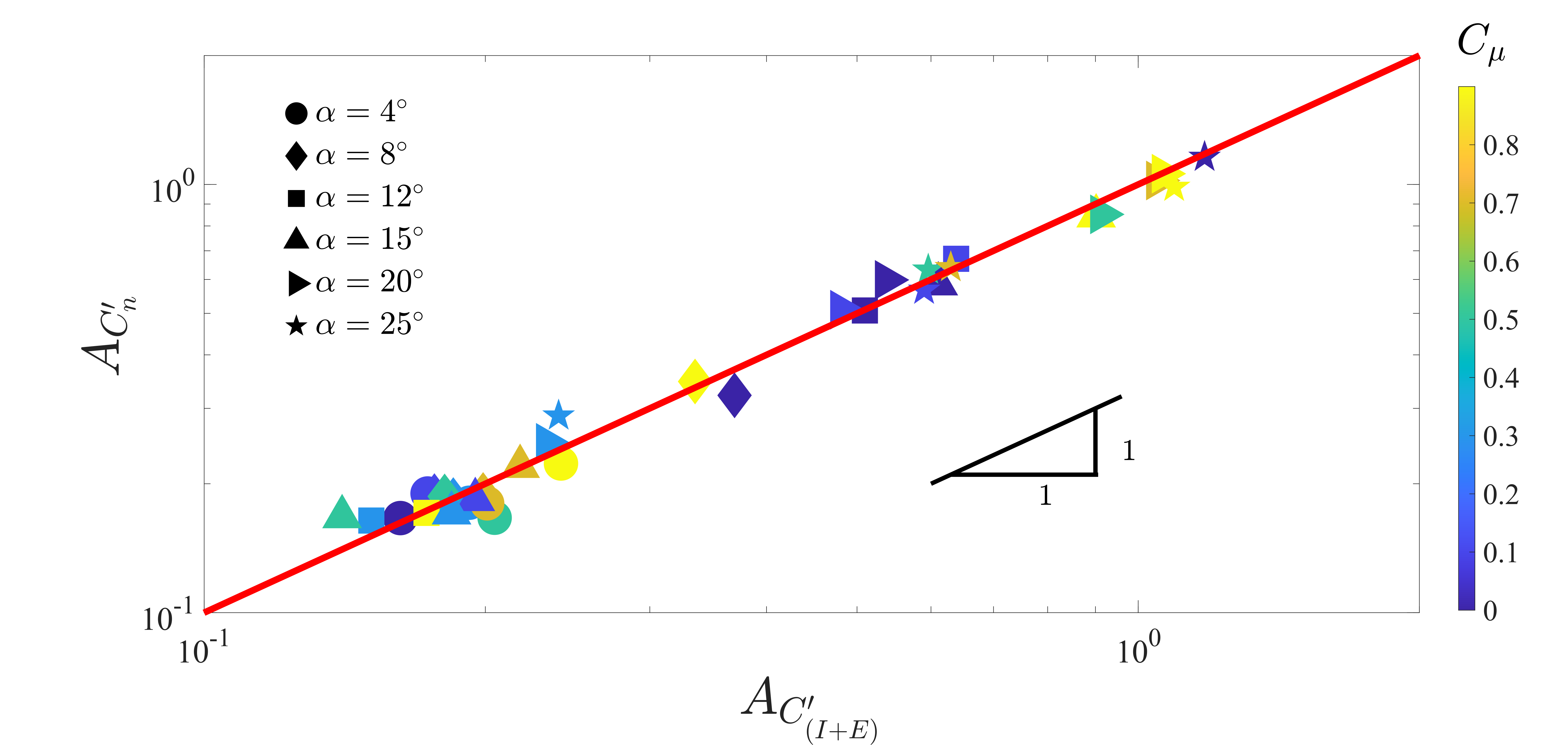}\label{fig:dis_web}}
	\caption{\label{fig:dis_we}Scaling relations for (a) the time-averaged maximum membrane deformation and Weber number and (b) nondimensional amplitudes of unsteady force and kinematics fluctuations. The coefficients of the fitting line in (a) is $c_0$=0.3178 and $c_1$=0.3212.}
\end{figure}

Two scaling relations presented in \cite{li2023unsteady} are utilized herein to characterize the connection between the aerodynamic performance and the membrane deformation from two aspects, namely the camber effect and the flow-induce vibration. In \reffig{fig:dis_we} \subref{fig:dis_wea}, we present the relation for the time-averaged membrane camber and the mean normal force represented by a nondimensional variable, the so called Weber number $We$. The scaling relation following the power law is given as
\begin{equation}
	\frac{\overline{\delta}_n^{max}}{c}=c_0 \ (We)^{c_1} = c_0 \ \left(\frac{C_n}{Ae}\right)^{c_1},
	\label{We} 
\end{equation}
where $c_0$ and $c_1$ are the fitting coefficients. The Weber number is the ratio between the normal force coefficient $C_n$ and the aeroelastic number $Ae$, which is given as
\begin{equation}
	We = \frac{F_n}{E^s h} = C_n \frac{\frac{1}{2} \rho^f U_{\infty}^2 c}{E^s h} = \frac{C_n}{Ae}
	\label{We2} 
\end{equation}

The labels in \reffig{fig:dis_we} \subref{fig:dis_wea} are colored by the momentum coefficient $C_{\mu}$ and different label shapes represent the angle of attack values. It can be observed from the plot that the data labels are reasonably collapsed onto the red curve indicated by \refeq{We}. The fitting coefficients are obtained from our previous study that investigates the membrane aeroelasticity in the parameter space of angle of attack and aeroelastic number. The good agreement of the data in the current study indicates that the aerodynamic performance still follows the power law of the wing camber even when high-momentum jet flows are injected into the coupled system. Besides, the label color gradually changes from blue to yellow to the right along the red curve. It indicates that a higher momentum coefficient can increase the membrane camber and produce a larger aerodynamic force. 

The scaling relation for aerodynamic force fluctuations and membrane vibrations is presented in \reffig{fig:dis_we} \subref{fig:dis_web}. The expression of this scaling relation is given as
\begin{equation}
	\left( 8\pi^2 m^* + \frac{16 \pi}{i}  \right) {St}^2 A^*_{\delta_n^{\prime}} + a^i + 4 Ae (\lambda -1) A^*_{\delta_n^{\prime}} =  A_{C_n^{\prime}}
	\label{unsteady_membrane13} 
\end{equation}
where $m^*$, $i$, $A^*_{\delta_n^{\prime}}$, $St$ and $\lambda$ denote the mass ratio, the structural mode number, the chord-normalized membrane displacement amplitude, the Strouhal number and the membrane stretch ratio, respectively. The left-hand side of \refeq{unsteady_membrane13} is an inertial-elastic combined force amplitude nondimensional number $A_{C_{(I+E)}^{\prime}}$. The detailed derivation of \refeq{unsteady_membrane13} can be found in our previous study \cite{li2023unsteady}. A similar scaling relation was proposed by Mathai et al. \cite{mathai2023shape} for a head-on placed membrane immersed in a uniform flow to characterize the relation between the drag coefficient fluctuation $A_{C_D^{\prime}}$ and the membrane vibration $A^*_{\delta_n^{\prime}}$, which is given as $A_{C_D^{\prime}} \approx 8 \pi^2 m^* {St}^2 A^*_{\delta_n^{\prime}}$. The forces produced by the inertial effect are considered in this relation and the elastic effect as well as the structural mode shape are neglected. It can be seen from \reffig{fig:dis_we} \subref{fig:dis_web} that all data is well collapsed onto a straight line in red color with slope 1. The novel scaling relation proposed in \cite{li2023unsteady} still performs very well even when the active jet flow control is applied to the coupled system. In \reffig{fig:dis_we} \subref{fig:dis_web}, the labels in the lower left corner are brightly colored and correspond to the parameter combinations for which the active jet flow control works. According to the physical meaning reflected by the scaling relation, we can infer that applying appropriate jet flows can effectively suppress aerodynamic force fluctuations by transferring momentum to the separation region, thereby suppressing flow-induced vibration.

%$\boldsymbol{F}(t)=\rho^f \frac{\text{d}}{\text{d}t} \left( \iint_{V_{body}} \boldsymbol{u}^f \text{d}x \text{d}z - \iint_{V_{\infty}} \boldsymbol{x} \times \boldsymbol{\omega} \text{d}x \text{d}z  \right)$

\subsection{Feedback loop and guidelines for jet-based flow control} \label{sec:section4.5}
\begin{figure}[H]
	\centering
	\includegraphics[width=1.0 \textwidth]{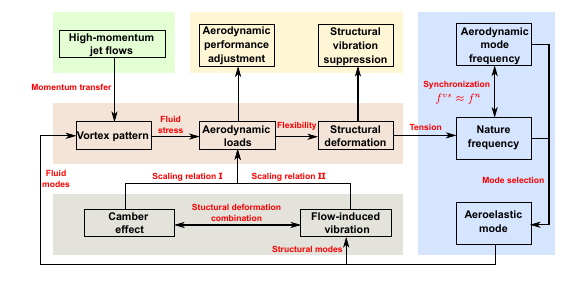}
	\caption{\label{fig:mechanism} Feedback loop of steady jet-based control mechanism in the coupled fluid-membrane system. The expressions of scaling relations I and II are given in \refeq{We} and \refeq{unsteady_membrane13}, respectively.}
\end{figure}

Based on the mechanisms discussed in \refse{sec:section4.4} and combined with a series of our previous studies \cite{li2022aeroelastic,li2020flow_accept,li2021high,li2023unsteady}, a unifying feedback loop that quantitatively connects the vortex pattern, the aerodynamic loads and the structural deformation on the basis of physical equations is suggested in \reffig{fig:mechanism}. This feedback loop clearly reveals the underlying mechanisms for the unsteady fluid flows coupling with the flexible membrane to drive structural deformation and produce aerodynamic loads. 

In this feedback loop, the vortex pattern directly influences the pressure distributions and governs the aerodynamic loads $\boldsymbol{F}(t)$ through a general formula derived from the momentum equation by Wu \cite{wu1981theory}, which is given as 
\begin{equation}
	\includegraphics[valign=c,width=0.5 \textwidth]{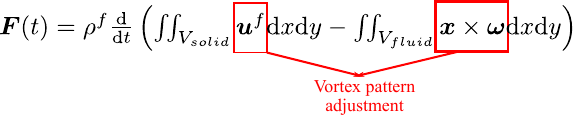}
	\label{suggestion1}
\end{equation}
where $V_{solid}$ and $V_{fluid}$ denote the membrane body volume and the total fluid volume including the body, respectively. Herein, $\boldsymbol{\omega}$ is the vorticity distribution at the position $\boldsymbol{x}$. The aerodynamic loads can be calculated only from the vorticity field without the need to compute the shear stress and the pressure on the membrane wing. On the one hand, the aerodynamic loads are coupled with the flexible membrane to drive structural deformation $\delta_n$ through the flexibility effect based on the structural governing equation, which is given as 
\begin{equation}
	\includegraphics[valign=c,width=0.3 \textwidth]{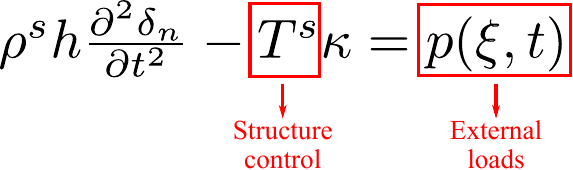}
	\label{suggestion2}
\end{equation}
where $\rho^s$ and $h$ are the material density and thickness, respectively. Herein, $T^s$, $\kappa$ and $p(\xi,t)$ denote the membrane tension, the curvature of the deformed membrane and the pressure distribution on the surface. The natural frequency of the flexible membrane is adjusted by the membrane tension and synchronized with the fluid mode frequency to excite the membrane instability through the aeroelastic mode selection. In turn, the selected fluid modes form particular vortex patterns and the selected structural modes excite particular flow-induced vibration. On the other hand, the structural deformation governs the aerodynamic loads from two aspects, namely the camber effect and the flow-induced vibration. These two factors involved in the structural deformation are directly connected with the aerodynamic force generation through two scaling relations in \refeq{We} and \refeq{unsteady_membrane13} presented in \refse{sec:section4.4}. So far, a complete feedback loop has been formed to reflect the fluid-membrane coupling mechanism when the flow-excited membrane instability occurs.

The proposed feedback loop provides straightforward guidance on designing optimal active flow control strategies from the physical level to achieve the required flight performance. It can be seen from \refeq{suggestion1} and \refeq{suggestion2} that the coupled fluid-membrane dynamics can be adjusted from three aspects, which are structure control by tuning membrane tension $T^s$ \cite{curet2014aerodynamic,bohnker2019control,bohnker2023integrated,buoso2015electro,buoso2017demand,buoso2017bat,he2023aerodynamics,huang2021fluid,huang2022energy}, external loads $p(\xi,t)$ \cite{tiomkin2022unsteady,sun2016nonlinear} and vortex pattern adjustment in the vorticity field $\boldsymbol{\omega}$, respectively. The first two active control strategies developed for flexible membranes have been widely explored over the past decades. Unlike these two strategies, based on the understanding of the coupling mechanism from the proposed feedback loop, we try to develop novel active control strategies from the viewpoint of vortex pattern adjustment and fluid transport. In this study, we demonstrate for the first time an active jet flow control strategy to achieve aerodynamic performance adjustment and vibration suppression in coupled fluid-membrane systems.

The results presented in this study verify that the active jet flow control has the capability to achieve aerodynamic performance adjustment and vibration suppression for flexible membrane wings. For membrane wings at relatively low angles of attack, the jet flows applied at the leading edge can be attached to the membrane surface and improve the momentum in the boundary layer due to the Coanda-like effect. When the momentum of the jet flows is high enough, the wing camber greatly increases, thereby improving the lift performance. Besides, the jets with high momentum injected into the boundary layer delay the flow separation and push the vortices away from the surface, leading to vibration suppression. As for membrane wings at high angles of attack in stall conditions, the strong adverse pressure gradient of the vortices prevents the jet flows released at the leading edge from being transported to the separation region. Current jet actuator placement strategies in this study do not work well for membrane wings that experience large flow separations at high angles of attack. 

Based on the fluid transport mechanism reflected by LCSs in \refse{sec:section4.4}, the jet actuator can be placed in the middle of the membrane and the jet direction should be close to the membrane surface. This placement strategy can keep the generated leading edge vortex which is beneficial to lift improvement, while eliminating the vortex shedding in the rear part to suppress membrane vibration. In addition to the application of jets to achieve membrane vibration suppression mentioned above, the jets can be also used to facilitate aeroelastic mode transition for selecting particular modes, enhance membrane vibration for energy harvesting and adjust aerodynamic performance under multiple flight mission conditions. The number of jet actuators, jet location, frequency and amplitude of unsteady jets should be further explored for the coupled fluid-membrane system. Some novel active flow control methods, like synthetic jet and co-flow jet, can be considered to be incorporated into the coupled fluid-membrane system to prompt the development of next-generation morphing aircraft.

\section{Conclusions} \label{sec:section5}
In this study, we demonstrated the application of active jet flow control for flexible membrane wings to achieve optimal flight performance. The coupled fluid-membrane dynamics were numerically simulated by a computational aeroelastic solver in the parameter space of the angle of attack and the momentum coefficient. The lift performance was continuously improved as the momentum coefficient increased for the flexible membrane within $\alpha < 12^\circ$, while it showed a downward trend and then grew up at high angles of attack. The maximum lift force was achieved at $\alpha=12^\circ$ with the maximum momentum coefficient. The current active jet flow placement strategy was not beneficial to the drag reduction and the lift-to-drag ratio performance. The membrane camber showed an overall positive correlation with the momentum coefficient. The lift fluctuation and the membrane vibration exhibited similar contour distributions, indicating an underlying connection. The active jet flow control showed its capability to suppress the aerodynamic force fluctuation and the membrane vibration for all momentum coefficients under $\alpha < 12^\circ$, while the control played a role for moderate momentum coefficients at high angles of attack near the stall condition. By comparing the flow features and the membrane responses at different momentum coefficients, we found that the high-momentum jet flows were transported to the boundary layer and got attached to the membrane surface due to the Coanda-like effect at low angles of attack, thereby reducing flow separation, improving lift performance and suppressing membrane vibration. On the contrary, the jet flows were enforced to keep away from the membrane surface due to the strong adverse pressure gradient in the separation region at high angles of attack, resulting in poor control effect on the aerodynamic performance and the membrane vibration. The effect of jet flow control on the drag variation was analyzed by decomposing the total drag into the viscous and pressure-induced components. The drag penalty was mainly contributed by the viscous drag component under active jet flow control. The pressure drag component was dominated in the total drag generation for high angles of attack.

We further investigated the vortex pattern evolution by extracting the dominant POD modes in the Eulerian view and examined the fluid transport process reflected by the attracting and repelling LCSs in the Lagrangian description. The high-momentum jet flows were transported to the wake and the vortex patterns were adjusted accordingly, thereby changing the aerodynamic loads governed by the momentum equation. The structural deformation was driven under the aerodynamic loads through the flexibility effect. The flexible membrane was coupled with the unsteady fluid flows to excite particular aeroelastic modes via frequency synchronization. Two scaling relations that connected the structural deformation and the aerodynamic force generation were verified even when the active flow control was applied to the coupled system. Finally, we suggested a unifying feedback loop to quantitatively link the vortex pattern evolution, the aerodynamic force generation and the membrane deformation and reveal the fluid-membrane coupling mechanism. Some guidelines for the application of active jet flow control in the coupled fluid-membrane system were summarized. The proposed feedback loop provided a useful technical roadmap to design optimal active jet flow control strategies for the flexible membrane wings. These findings can facilitate the development of next-generation morphing aircraft incorporating active flow control techniques and enhance flight adaptation to complex time-varying environments.

\section*{Acknowledgements}
The first author would like to acknowledge the support from the National Natural Science Foundation of China (NSFC) (Grant Number 12202362), the China Postdoctoral Science Foundation (Grant Number 2023M732798) and the High Performance Computing (HPC) Platform in Xi'an Jiaotong University. This work is also funded by the open fund in the State Key Laboratory for Manufacturing Systems Engineering of Xi'an Jiaotong University (Grant Number sklms2023014). This work is also supported by the Key Research and Development Program of Shaanxi Province (Grant Number 2021ZDLGY12-04), the National Natural Science Foundation of China (Grant Number 52075433).  The second author would like to acknowledge the support
from the University of British Columbia and the Natural Sciences and Engineering Research Council of Canada.

\section*{Declaration of Competing Interest}
The authors declare that they have no known competing financial interests or personal relationships that could have appeared to
influence the work reported in this paper.

\section*{Data availability}
Data availability.

\section*{Appendix A: Mesh convergence study}
\setcounter{equation}{0}
\setcounter{figure}{0}
\setcounter{table}{0}
\renewcommand{\theequation}{A.\arabic{equation}}
\renewcommand{\thefigure}{A.\arabic{figure}}
\renewcommand{\thetable}{A.\arabic{table}}

To choose an appropriate mesh resolution in the numerical simulation, we perform a mesh convergence study before proceeding to the systematical investigation of the steady wake jet effect. We design three types of meshes, namely M1, M2 and M3, to discretize the fluid and structural domains. A typical state at $\alpha$=$20^\circ$ with $C_{\mu}$=0.5 is selected to conduct the mesh convergence study. Large-scale separation flows are observed at this angle of attack and the lift can be improved at the chosen moderate momentum coefficient.

\begin{figure}[H]
	\centering
	\subfloat[]{\includegraphics[width=0.49 \textwidth]{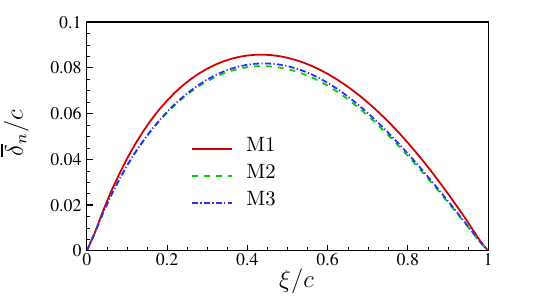}\label{fig:convergencea}}
	\subfloat[]{\includegraphics[width=0.49 \textwidth]{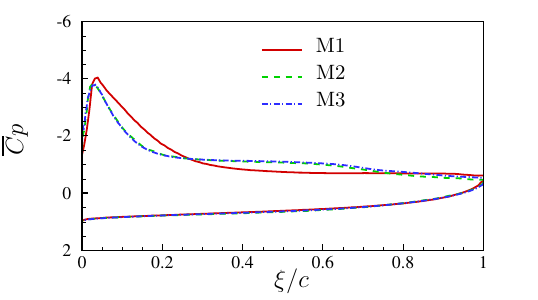}\label{fig:convergenceb}}
	\caption{\label{fig:convergence}Flow past a flexible membrane with steady jet flow at $Re$=$2500$ with an angle of attack of $\alpha$=$20^\circ$ and a momentum coefficient of $C_{\mu}$=$0.5$: comparison of time-averaged (a) membrane displacement and (b) pressure coefficient.}
\end{figure}

A comparison of the aerodynamic characteristics, membrane displacement and Strouhal number for three different meshes is summarized in \reftab{tab:mesh convergence}. Mesh M3 is selected as the reference to calculate the percentage differences. It can be seen from \reftab{tab:mesh convergence} that the percentage differences between M1 and M3 are larger than $5\%$. As the mesh is refined to M2, the maximum percentage difference is noticed as $2.44\%$ for the membrane displacement at the middle chord. \refFig{fig:convergence} further compares the time-averaged membrane displacement and pressure distribution along the membrane surface between the three different meshes. The membrane responses between M2 and M3 show very small differences. Thus, the mesh resolution of M2 is adequate to capture the coupled fluid-membrane dynamics and is selected as the reference mesh in further numerical studies.

\begin{table}[H]
	\centering
	\caption{Aerodynamic characteristics, membrane displacement at the middle chord and non-dimensional vortex shedding frequency for mesh convergence of a flexible membrane wing at $Re$=$2500$ with an angle of attack of $\alpha$=$20^\circ$ and a momentum coefficient of $C_{\mu}$=$0.5$. The percentage differences are calculated by using M3 results as the reference.}{\label{tab:mesh convergence}}
	\begin{tabular}{c c c c c c c}
		\toprule  
		Mesh & $\overline{C}_L$ & ${C_L^{\prime}}^{rms}$ & $\overline{C}_D$ & ${C_D^{\prime}}^{rms}$ & $\overline{\delta}^{middle}_n/c$ & $f^{vs}c/U_{\infty}$ \\ [3pt]
		\midrule 
		M1 & 1.6340 (-5.98\%) & 0.5750 (-12.33\%) & 0.4970 (-7.71\%) & 0.2218 (-23.96\%)  & 0.0864 (5.24\%)& 0.5774 (-9.11\%) \\  
		M2 & 1.7372 (-0.04\%) & 0.6704 (2.21\%) & 0.5359 (-0.48\%) & 0.2861 (-1.92\%)  & 0.0801 (-2.44\%) & 0.6353 (0\%)  \\  
		M3 & 1.7379 & 0.6559 & 0.5385 & 0.2917 & 0.0821  & 0.6353 \\  
		\bottomrule 
	\end{tabular}
\end{table}

\section*{APPENDIX B: Comparative study and validation} \label{sec:section8}
\setcounter{equation}{0}
\setcounter{figure}{0}
\setcounter{table}{0}
\renewcommand{\theequation}{B.\arabic{equation}}
\renewcommand{\thefigure}{B.\arabic{figure}}
\renewcommand{\thetable}{B.\arabic{table}}
A comparative study of the aerodynamic characteristics is performed for a two-dimensional flexible membrane in \reffig{fig:comparative}. The membrane model is the same as the model established in \refse{sec:section3} except for the jet flow boundary at the leading edge. A non-slip wall boundary is imposed instead of the jet boundary. The coupled fluid-membrane dynamics is simulated from $\alpha=4^\circ$ to $20^\circ$. The Reynolds number, the mass ratio and the aeroelastic number employed in the simulation are the same as the model set up in \refse{sec:section3}. A comparison of the time-averaged aerodynamic forces, the deformed membrane profile and the pressure distribution is performed in \reffig{fig:comparative}. A good agreement is observed between our high-fidelity numerical results and the data obtained from Sun et al. \cite{sun2017effect}.

\begin{figure}[H]
	\centering
	\subfloat[]{\includegraphics[width=0.49 \textwidth]{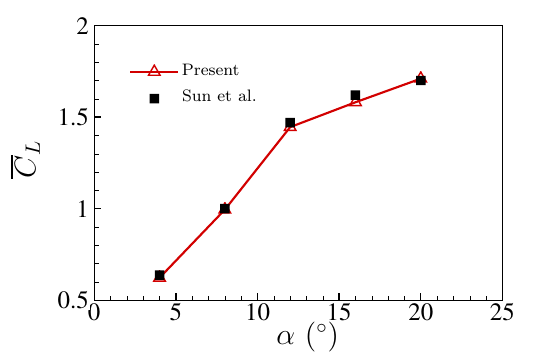}\label{fig:comparativea}}
	\subfloat[]{\includegraphics[width=0.49 \textwidth]{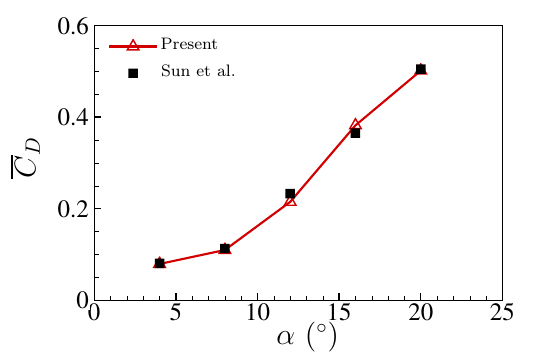}\label{fig:comparativeb}}
	\\
	\subfloat[]{\includegraphics[width=0.49 \textwidth]{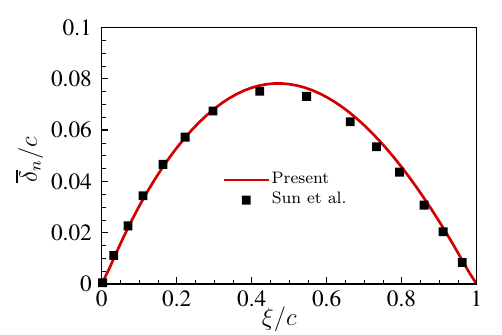}\label{fig:comparativec}}
	\subfloat[]{\includegraphics[width=0.49 \textwidth]{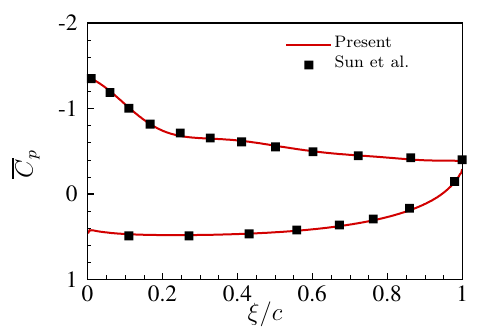}\label{fig:comparatived}}
	\caption{\label{fig:comparative}Comparison of time-averaged (a) lift coefficient, (b) drag coefficient, (c) membrane deformation under $\alpha=8^\circ$ and (d) pressure coefficient distribution under $\alpha=8^\circ$ for a two-dimensional flexible membrane at $Re$=2500 between the numerical results from our aeroelastic solver and Sun et al. \cite{sun2017effect}.}
\end{figure}

\begin{figure}[H]
	\centering
	\subfloat[]{\includegraphics[width=0.49 \textwidth]{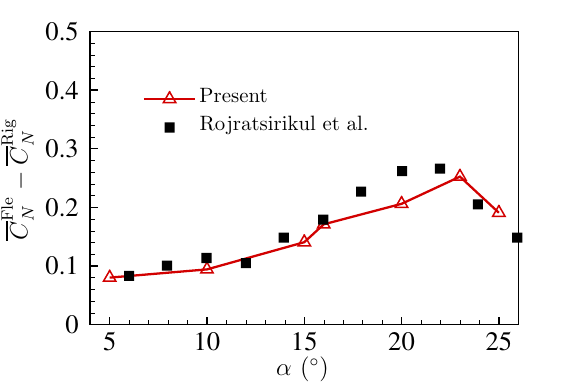}\label{fig:validationa}}
	\subfloat[]{\includegraphics[width=0.49 \textwidth]{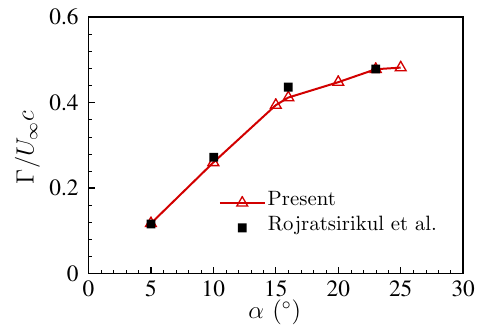}\label{fig:validationb}}
	\\
	\subfloat[]{\includegraphics[width=0.49 \textwidth]{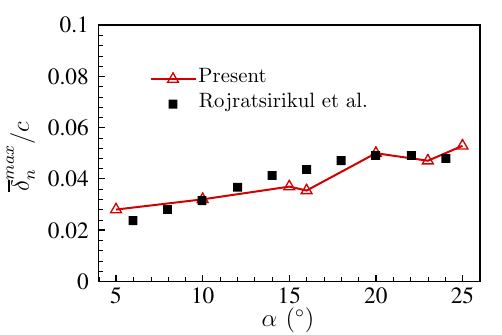}\label{fig:validationc}}
	\subfloat[]{\includegraphics[width=0.49 \textwidth]{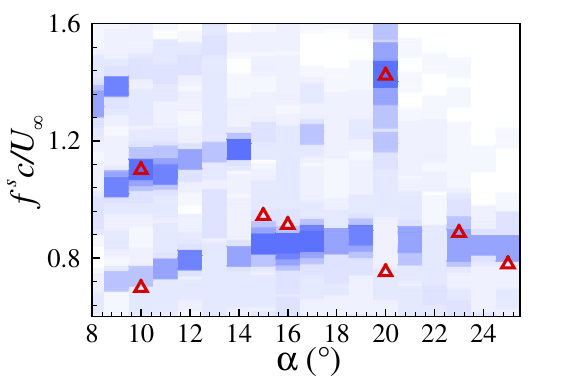}\label{fig:validationd}}
	\caption{\label{fig:validation} Comparison of (a) time-averaged normal force coefficient variation relative to rigid wing, (b) normalized circulation of the wingtip vortices, (c) time-averaged maximum membrane deformation and (d) membrane vibration frequency spectra between the numerical results from our aeroelastic solver and experimental data performed by Rojratsirikul et al. \cite{rojratsirikul2010unsteady,rojratsirikul2011flow}.}
\end{figure}

We further validate the coupled fluid-membrane dynamics of a three-dimensional membrane model presented in experimental studies \cite{rojratsirikul2010unsteady,rojratsirikul2011flow} by using our high-fidelity aeroelastic solver. The wing has a chord length of 68.75 mm and an aspect ratio of 2. The geometry details can be found in Li et al. \cite{li2020flow_accept}. The wing with Young's modulus of 2.2 MPa and a structural density of 1000 $\text{kg} \cdot \text{m}^{-3}$ is placed in a uniform fluid flow with $5$ m/s. The aerodynamic performance, the flow features, and the structural responses are calculated by the high-fidelity aeroelastic solver and compared with experimental data in \reffig{fig:validation} for validation purposes. The coupled fluid-membrane dynamics are reasonably predicted by our aeroelastic solver.

%\newpage
\section*{References}

\bibliography{referenceBib}

\end{document}